\RequirePackage{fix-cm}
\documentclass[smallextended]{svjour3}       % onecolumn (second format)
\smartqed  % flush right qed marks, e.g. at end of proof
\usepackage{graphicx}
\usepackage{mathptmx}      % use Times fonts if available on your TeX system

\journalname{Experimental Astronomy}

\usepackage{times,epsfig,pdfpages,fancybox,rotating}

\usepackage{pifont,url,ulem}

\usepackage{natbib}       \bibpunct{[}{]}{;}{n}{}{,}

\newcommand{\arcsec}{\ensuremath{''}\,}

\begin{document}

\title{LEMUR: Large European Module for solar Ultraviolet Research}
\subtitle{European contribution to JAXA's Solar-C mission}

%\titlerunning{LEMUR: European contribution to Solar-C}        % if too long for running head

\author{Luca Teriaca \and Vincenzo Andretta \and Fr\'{e}d\'{e}ric Auch\`{e}re \and 
Charles M. Brown \and Eric Buchlin \and Gianna Cauzzi \and
J. Len Culhane \and Werner Curdt \and Joseph M. Davila \and 
Giulio Del Zanna \and George A. Doschek \and Silvano Fineschi \and
Andrzej Fludra \and Peter T. Gallagher \and Lucie Green \and
Louise K. Harra \and Shinsuke Imada \and Davina Innes \and 
Bernhard Kliem \and Clarence Korendyke \and John T. Mariska \and 
Valentin Mart\'{i}nez-Pillet \and Susanna Parenti \and 
Spiros Patsourakos \and Hardi Peter \and Luca Poletto
\and Rob Rutten \and Udo Sch\"{u}hle \and Martin Siemer \and 
Toshifumi Shimizu \and Hector Socas-Navarro \and 
Sami K. Solanki \and Daniele Spadaro \and Javier Trujillo-Bueno \and
Saku Tsuneta \and Santiago Vargas Dominguez \and
Jean-Claude Vial \and Robert Walsh \and Harry P. Warren \and
Thomas Wiegelmann \and Berend Winter \and Peter Young
}

%\authorrunning{Teriaca et al.} % if too long for running head

\institute{
L. Teriaca, W. Curdt, D. Innes, H. Peter, U. Sch\"{u}hle, S. K. Solanki, T. Wiegelmann \at
Max-Planck-Institut f\"ur Sonnensystemforschung\\
Max Planck str. 2, 37191 Katlenburg-Lindau, Germany\\
\email{teriaca@mps.mpg.de}
\and
V. Andretta \at
INAF - Osservatorio Astronomico di Capodimonte, Salita Moiariello 16, 80131 Napoli, Italy     
\and
F. Auch\`{e}re, E. Buchlin, J.-C. Vial \at
Institut d'Astrophysique Spatiale, B\^{a}timent 121, Universit\'{e} Paris-Sud 11, 91405 Orsay, France
\and
C.M. Brown, G.A. Doschek, C. Korendyke, J.T. Mariska, H.P. Warren \at
Space Science Division, Naval Research Laboratory, Washington DC,
20375-5320, USA
\and
G. Cauzzi \at
INAF - Osservatorio Astrofisico di Arcetri, 50125 Florence, Italy
\and
J.L. Culhane, L. Green, L.K. Harra, S. Vargas Dominguez, B. Winter \at
UCL - Mullard Space Science Laboratory, Holmbury St. Mary, Dorking, Surrey RH5 6NT, UK
\and
J.M. Davila \at
NASA - Goddard Space Flight Center, Greenbelt, MD. USA
\and
G. Del Zanna \at
University of Cambridge, UK
\and
S. Fineschi \at
INAF - Osservatorio Astronimico di Torino, 20 Strada Osservatorio, Pino Torinese, Italy
\and
A. Fludra \at
STFC Rutherford Appleton laboratory, UK
\and
P.T. Gallagher \at
School of Physics, Trinity College Dublin, Dublin 2, Ireland
\and
S. Imada, T. Shimizu \at
Institute of Space and Astronautical Science, JAXA, 3-1-1 Yoshinodai, Chuo-ku, Sagamihara, 
Kanagawa 252-5210, Japan
\and
B. Kliem \at
Institute of Physics and Astronomy, University of Potsdam, Karl-Liebknecht-Str. 24-25, 
Potsdam 14476, Germany
\and
V. Mart\'{i}nez-Pillet, H. Socas-Navarro, J. Trujillo-Bueno\at
Instituto de Astrof\'{i}sica de Canarias, 38205 La Laguna, Tenerife, Spain
\and
S. Parenti \at
Royal Observatory of Belgium, Belgium
\and
S. Patsourakos \at
University of Ioannina, Department of  Physics-Astrogeophysics Section, GR 451 10 Ioannina, Greece
\and
L. Poletto \at
CNR - Institute of Photonics and Nanotechnologies, Padua, Italy
\and
R. Rutten \at
Sterrekundig Instituut Utrecht, The Netherlands
\and
M. Siemer \at
DLR - Institute of Space Systems, Bremen, Germany
\and
D. Spadaro \at
INAF - Osservatorio Astrofisico di Catania, Via S. Sofia 78, 95123 Catania, Italy
\and
S. Tsuneta \at
National Astronomical Observatory of Japan, 2-21-1 Osawa, Mitaka, Tokyo 181-8588, Japan
\and
R. Walsh \at
University of Central Lancashire, UK
\and
P. Young \at
George Mason University, USA
}

\date{Received: date / Accepted: date}
% The correct dates will be entered by the editor

\maketitle

\begin{abstract}
The solar outer atmosphere is an extremely dynamic
environment characterized by the continuous
interplay between the plasma and the magnetic field
that generates and permeates it.  Such
interactions play a fundamental role in hugely 
diverse astrophysical systems, but occur at scales 
that cannot be studied outside the solar system.

Understanding this complex system requires 
concerted, simultaneous solar observations from the 
visible to the vacuum ultraviolet (VUV) and soft X-rays, 
at high spatial resolution (between 0.1\arcsec and 0.3\arcsec), at 
high temporal resolution (on the order of 10~s, i.e., the 
time scale of chromospheric dynamics), with a wide 
temperature coverage (0.01~MK to 20~MK, from the 
chromosphere to the flaring corona), and the 
capability of measuring magnetic fields through 
spectropolarimetry at visible and
near-infrared wavelengths.
Simultaneous spectroscopic measurements sampling 
the entire temperature range are particularly important.

These requirements are fulfilled by the Japanese Solar-C 
mission (Plan B), composed of a spacecraft
in a geosynchronous orbit with a payload providing a
significant improvement of imaging and
spectropolarimetric capabilities in the UV, visible, and
near-infrared with respect to what is available today
and foreseen in the near future. 

The Large European Module for solar Ultraviolet Research 
(LEMUR), described in this paper, is a large VUV telescope 
feeding a scientific payload of high-resolution imaging 
spectrographs and cameras. LEMUR consists of two major 
components: a VUV solar telescope  with a 30~cm diameter
mirror and a focal length of 3.6~m, and a focal-plane 
package composed of VUV spectrometers
covering six carefully chosen wavelength ranges between 
170~\AA\ and 1270~\AA. The LEMUR slit covers 
280$\arcsec$ on the Sun with 0.14$\arcsec$ per pixel 
sampling. 
In addition, LEMUR is capable of measuring
mass flows velocities (line shifts) down to
2~km~s$^{-1}$ or better.

LEMUR has been proposed to ESA as the European contribution to the 
Solar C mission.

\end{abstract}

\section{Introduction}
\label{intro}
For life on Earth, the Sun is the most important 
astrophysical object in the universe, and it holds a
special place in the human imagination.
Solar eclipses and the elusive appearance of the ghostlike corona
have been a source of fascination since pre-historic
times.  Unfortunately, the impossibility of {\it in-situ} measurements
at or near the solar surface and the complexity of the solar
atmosphere make it difficult to study. Solar observations during the
past several decades have yielded a wealth of new information and deep
insights into the physical processes that occur in the solar
atmosphere. Yet many questions remain unresolved. These include the
heating of the solar chromosphere, the layer above the solar surface
in which the temperature starts to increase with height, and the million
degree corona (compared with the 6000 degree solar surface), the onset
of coronal mass ejections -- huge clouds of hot gas flung from the Sun
at speeds of several 1000~km\,s$^{-1}$ -- and the origin of the
solar wind, a constant outflow of coronal gas that fills the
heliosphere and extends the solar atmosphere out to 100~AU or more.
Since these processes generate and simultaneously perturb the solar
heliosphere, their understanding is central for answering the Cosmic
Vision question: {\it How does the solar system work\/}?  Moreover, the
solar wind carries turbulent magnetic field out to the edge of the
solar system where it drastically reduces the flux of 
incoming cosmic rays.  These high-energy particles are harmful to life
forms, especially highly evolved life forms.  
Understanding the solar magnetic system, its variability and severe
modifications in large solar eruptions is therefore pivotal to
answering the Cosmic Vision question: {\it What are the
conditions for planet formation and the emergence of life\/}?  It is
particularly relevant to the sub-topic {\it Life and habitability in
the solar system}.

In almost all cases the primary impediment to progress in answering
these questions is our inability to resolve and follow the small-scale
structures produced by the processes that energize the solar
atmosphere. Since these structures are intimately connected with the
solar magnetic field, it is largely our inability to track the magnetic field
from the photosphere into the corona that impedes progress.  Many of
the physical processes that produce activity in the solar atmosphere
involve the creation, structuring, and dissipation of magnetic fields
in a plasma, fundamental processes in all of astrophysics.  Only on
the Sun, however, do we have the possibility at least in principle to
resolve the scales over which these processes take place, which makes
studying them on the Sun of particular relevance for astrophysics as a whole.  

Section~\ref{sci-obj-req-solarc} describes the scientific goals of the Solar C 
mission and the payload devised to achieve them. The specific scientific 
motivations of LEMUR and the derived technical requirements are discussed in
Section~\ref{sci-obj-req-lemur}. The mission profile is briefly discussed in
Section~\ref{sec-mission-profile} while a detailed technical description of LEMUR 
is given in Section~\ref{sec-mod-payload}. Finally, Section~\ref{sec-concl}
summarizes the paper.

\section{Scientific Motivation for the Japanese Solar-C Mission}
\label{sci-obj-req-solarc}

During the last two decades, the availability of VUV (vacuum
ultraviolet: 150 to 2000 \AA) and X-ray telescopes and spectrometers
with ever-increasing capabilities have greatly enriched our knowledge
of the outer solar atmosphere. The Yohkoh mission provided detailed 
observations of coronal loops, jets, coronal arcade formation, and long-duration
cusp-shaped flares.  The SOHO mission gave us a global view of
solar transition region dynamics and structure, and made a giant leap in our
observations and understanding of coronal mass ejections.  Hinode data
are providing new knowledge on photospheric and chromospheric dynamics
and energy transport mechanisms, detailed temperature and density
structures of active regions and their related dynamics, and are
allowing tests of active region heating models such as the nanoflare
model.  However, key problems remain still unresolved. They include 
determining what role magnetic reconnection plays in
producing solar flares (impulsive events that produce intense
extreme-ultraviolet, X-ray, and sometimes gamma radiation) and coronal
mass ejections, how particles are accelerated in the solar atmosphere,
how the solar wind is energized and accelerated, and how small-scale
sites of energy release determine the large-scale structures of solar
activity.

During the past year solar physicists in Japan, the United States, and
Europe have been studying the possibility of a follow-on mission to
the highly successful Solar-A (Yohkoh) and Solar-B (Hinode) programs
that would make major steps forward in improving our understanding 
of the solar atmosphere. 

Solar-C will build on questions raised by but not answered by Hinode, 
in particular those regarding the dynamical coupling
between the various components of the solar atmosphere, considered as
an ``integrated system''.  Indeed, until now, solar physics space
research has mainly concentrated on selected regions of the solar atmosphere,
such as the corona or photosphere, and considerable progress has been
made in observing structures and dynamics in these regions.
Nevertheless, important Hinode results such as the discoveries of
chromospheric microjets and ``type-II'' spicules suggest that what is
required is the capability of measuring energy release over the entire
range of spatial, spectral, and temporal scales observed in different
atmospheric regions.  These scales often differ markedly in the
chromosphere, transition region, and corona.  Currently, even the
combination of data from several existing missions do not produce
adequate scale coverage.  Combinations of data from different missions
also tend to be restricted by difficulties in data co-registration and
simultaneity.

The unifying concept behind Solar-C is thus the need to understand how
the solar magnetic field drives the flow of mass and energy from the
thermally dominated lower solar atmosphere into the magnetically
dominated corona and solar wind acceleration region as a coupled
system. Understanding how this flow takes place requires detailed
knowledge of the magnetic field not only in the photosphere but also
above the $\beta=1$ surface (where $\beta$ is the 
kinetic gas to magnetic pressure ratio), in the chromosphere. 
Here, the proposed instrumentation for
Solar-C will make revolutionary new measurements of the magnetic field
with far better stability, polarimetric accuracy and temporal coverage
than is possible from ground-based telescopes
\citep[e.g.,][]{2010mcia.conf..118T}.  %Trujillo Bueno, review%
Besides providing accurate and proper 
boundary conditions for extrapolating the field into the corona,
knowledge of the chromospheric field also allows tracing the release
of free energy of the magnetic field and its effects on the momentum
and energy balance of the chromosphere and corona.
Since the forcing of the field implies that much of the magnetic energy
release occurs well above the photosphere, such studies are almost 
impossible using only photospheric fields. This, along with
all-encompassing plasma diagnostic capabilities and with the highest
resolution solar observations ever made at all temperatures present
throughout the atmosphere, will produce
fundamental advances in our understanding of how field and plasma
processes form structures, how magnetic energy is released through
instabilities, reconnection, wave dissipation, and particle
acceleration, and what role the magnetic fields play in
heating and accelerating plasma in the corona and the solar wind.
Solar-C will be the first mission to reveal the spatial
and temporal scales over which energy release in the solar atmosphere
is aggregated into observable structures.

Solar-C will address the following key questions, that form the mission science goals:
\begin{enumerate}  \vspace*{0.1ex} \itemsep=0ex 
  \it

\item How are elementary atmospheric structures created and
  how do they evolve in each temperature domain of the atmosphere?

\item How is energy transported through small elementary
  structures into the large-scale corona and how does it drive the
  solar wind?

\item How is magnetic energy dissipated in astrophysical
  plasmas?

\item How do small-scale physical processes initiate
  large-scale dynamic phenomena creating space weather?

\end{enumerate}

These key unresolved science questions can only be addressed with an
instrument suite with broad temperature coverage so that emission from
the photosphere, chromosphere, transition region, and corona can be
observed simultaneously. Another critical requirement for this mission
is high resolution.  High spatial resolution is necessary to reveal how 
the filamentary magnetic field in the photosphere is
relaxed from a forced to an almost force-free state at chromospheric
heights. High temporal resolution is necessary to capture the dynamics
of the chromosphere and transition region as they connect to the
corona, and high spectral resolution is required to provide
diagnostics of the magnetic field and energy flow through detailed
measurements of line profiles. 
These observational requirements can be met by a spacecraft placed 
in a geosynchronous orbit with a payload of three 
state-of-the-art instruments:

\begin{itemize} \itemsep=0ex \vspace{-0.5ex}

\item a Solar UltraViolet, Visible, and Infrared Telescope (SUVIT);

\item an X-ray or extreme-ultraviolet imaging telescope (XIT);

\item a vacuum ultraviolet (150\,\AA\ to 2000\,\AA) high-throughput
  spectroscopic telescope (proposed here with acronym LEMUR).

\end{itemize} \vspace{-1ex}

These instruments and their science cases have 
been studied in detail by the JAXA Solar-C 
International Working Group, with significant
contributions by scientists from European countries 
and the US.
Detailed descriptions are given in the Japanese Draft Interim
Report\footnote{SOLAR-C Working Group, 2011, Interim Report 
on the SOLAR-C Mission Concept.\\
\url{http://hinode.nao.ac.jp/SOLAR-C/archive_e.html}}.
The basic characteristics of
each instrument are summarized in Table~\ref{table:solarc-payload}.

%%%%%%%%%%%%%%%%%%%%%%%%%%%%%%
\begin{table*}[h]
\begin{center}
\caption[]{\label{table:solarc-payload}
           Solar-C instrumentation payload summary}
\begin{tabular}{ll}
\hline
Instrument characteristic & Nominal value \\
\hline  & \mbox{}\\[-1.5ex]
\multicolumn{2}{c}{LEMUR (Large European Module for solar
  Ultraviolet Research, proposed European contribution)}\\[0.5ex]
Telescope            & 30~cm diameter, off-axis paraboloid, f/12 \\
Focal plane package  & VUV single-grating spectrographs with visible 
                       light or \\
                     & UV slit imaging assembly\\
Wavelength coverage  & 170--210 \AA, 695--815 \AA, 965--1085 \AA, 
                       1150--1270 \AA\ \\
                     & 2$^{\rm nd}$ order 482--542 \AA, 575--635 \AA  \\  
Instantaneous field of view & 0.28$\arcsec\times$280\arcsec (300\arcsec$\times$280\arcsec by rastering) \\
Spatial, temporal resolution & 0.28\arcsec$^{1}$, (0.14\arcsec/pixel), down to 0.5~s (sit and stare)  \\
Spectral resolution             & $17000 \leq \lambda/\Delta\lambda \leq 32000$ \\
\hline & \mbox{}\\[-1.5ex]
\multicolumn{2}{c}{SUVIT (Solar UltraViolet, Visible, and Infrared 
  Telescope)}\\[0.5ex]
Telescope	     &  \O\ 1.5 m (nominal), reflective, Gregorian telescope (f/9.54)\\
Focal plane package  & Imaging and spectroscopic instrumentation, polarimetry \\
Wavelength coverage  & He\,I 10830\AA, D3 5876\AA, Ca\,II 8542\AA, 8498\AA, 
                       Fe\,I 8538\AA, \\
                     & Si\,I 10827\AA, Mg\,II h/k 2800\AA  \\
Spatial, temporal resolution$^2$ 
              & $\approx$0.1\arcsec\, (0.05\arcsec/pixel), 1~s\\
Field of view$^2$  &  $\approx$200\arcsec$\times$200\arcsec\\
\hline & \mbox{}\\[-1.5ex]
\multicolumn{2}{c}{XIT (Extreme ultraviolet Imaging Telescope, 
      normal incidence)$^{3}$}\\[0.5ex]
Telescope type	      & Normal incidence EUV multilayer telescope, 
                        \O\ 32 cm \\
Wavelength coverage   & Multi-layer narrow bands centered 
                        at 94, 171, 195, 211, 335\AA \\
                      & one UV band similar to SDO and TRACE 1600--1700\AA  \\
Spatial and temporal resolution 
                      &	0.2\arcsec with 0.1\arcsec/pixel, $<$ 10 s \\
Field of view	      & 400\arcsec$\times$400\arcsec \\
\hline & \mbox{}\\[-3.5ex]
\end{tabular}
\end{center}
\begin{small}
\mbox{}$^{1}$Based on preliminary error budget for the 170 to 210~\AA\, 
channel, a resolution of 0.14 to 0.20\arcsec\, (67\%\, encircled energy) 
appears achievable.\\
\mbox{}$^{2}$ Values refer to Narrow-band filtergrams at the shortest 
wavelengths. More complete and detailed information can be found in the
Interim Report on the SOLAR-C Mission Concept.\\
\mbox{}$^{3}$A 1\arcsec\ grazing incidence telescope 
with 0.5\arcsec/pixel plate scale is also being considered.  
This telescope would include a photon counting detector 
with energy discrimination capability for spectroscopy.
\end{small}
\end{table*}
%%%%%%%%%%%%%%%%%%%%%%%%%%%%%%%%%%%%%

\begin{figure}[t!]
\includegraphics[angle=270,width=120mm]{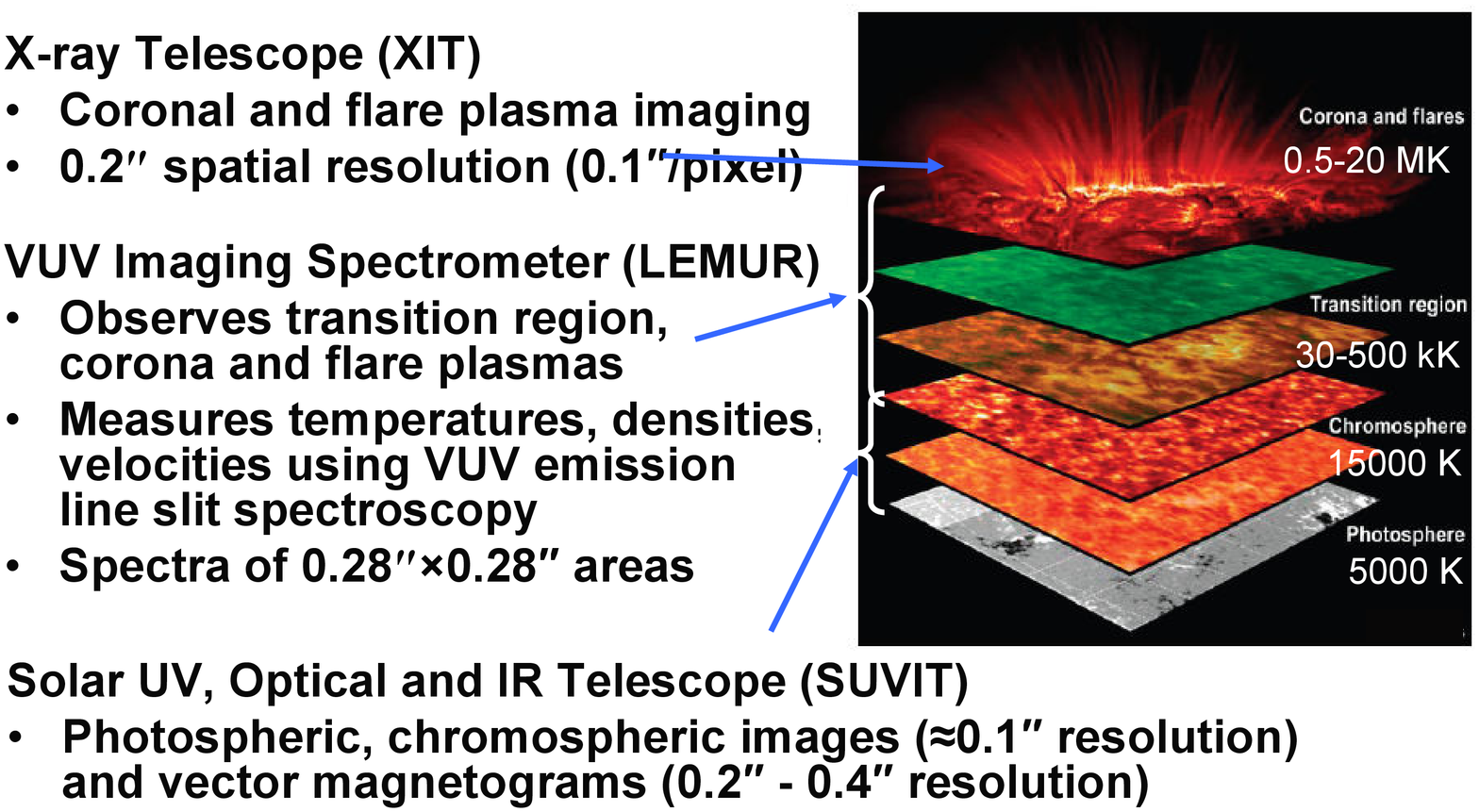} %%PS, PDF
\vskip -0.5cm
\caption[]{\label{Solar-C-schematic} The Solar-C observation
  space.  All regions of the solar atmosphere are observed with
  simultaneous and matched high spatial and spectral resolution.
  All figures are in color in the online version.}
\end{figure}

These instruments will provide overlapping
temperature coverage and complementary capabilities such as
spectroscopy and narrow wavelength band imaging, high cadence and
throughput, and will feature fields-of-view large enough
to properly study active regions. 
A pictorial representation of the overall Solar-C observational domain 
is shown in Figure~\ref{Solar-C-schematic}.
%%%%%%%%%%%%%%%%%%%%%%%%%%%%%
%%%%%%%%%%%%%%%%%%%%%%%%%%%%%

Solar-C will be highly synergistic with, and complementary to, the
forthcoming Solar Orbiter and Solar Probe+ 
missions\footnote{currently foreseen to be launched in 2017 and no 
later than 2018, respectively} of ESA and NASA.
Besides stereoscopic observations with Solar Orbiter, in-situ 
composition measurements made from Orbiter and Probe+ will be traced 
to their solar origin by Solar-C observations. 
Solar-C will greatly extend the important new information on the
chromosphere and transition region that will be provided by
the up-coming NASA IRIS mission. In fact, the latter does not have 
adequate transition region and coronal coverage, does not measure 
chromospheric magnetic fields, and will observe the chromosphere
with a lower spatial resolution than the Solar-C optical telescope.
Finally, Solar-C will also occur at a time when
major ground facilities such as the European Solar Telescope (EST) and
the Advanced Technology Solar Telescope (ATST) will become
operational, allowing highly complementary observations such as a
direct measurement of the coronal magnetic field at the limb.

Solar-C will have the ability to determine what energizes and
maintains the plasma in the complex magnetic environment of the solar
atmosphere.  It will be able to measure the signatures of all major
processes that form structures and release energy such as magnetic
reconnection and wave dissipation.  Theoretical work tells us that the
smallest spatial scales over which important physical processes such
as magnetic reconnection, wave dissipation, and particle acceleration
occur may be only meters in size, far smaller than will be resolvable
with solar telescopes for the foreseeable future. Solar observations,
however, suggest that these physical processes are coherent over much
larger spatial scales. Chromospheric jets observed with Hinode, for
example, have spatial scales of about 0.2\arcsec~to 0.3\arcsec\ (about 140 to
210~km at the Sun). Similarly, observations of transition region and
coronal structures taken with $>$~1\arcsec\ resolution instruments
indicate that at least
10\%\ of the observed area is filled with emitting
plasma, again suggesting structures with spatial scales that can be 
observed by the next generation of solar instruments.

It is widely believed that magnetic reconnection plays a central role
in many solar phenomena, such as flux emergence and cancellation in
the photosphere, chromospheric and coronal jets, and coronal eruptions
(flares and coronal mass ejections). The physics of magnetic
reconnection, however, is still poorly understood. To further progress,
the plasma conditions
in and around the current sheet must be explored in detail. 
This motivates the selection of multiple wavelength ranges so that the
brightest emission lines at critical temperatures can be observed
spectroscopically with high spatial resolution. Another aim for
Solar-C is to observe the signature of reconnection in the lower
atmosphere in order to explore this process in partially ionized
plasma, which is highly relevant in many astrophysical settings, for
example in star-forming regions.  This can also shed light on the
surprisingly short time scales of chromospheric energy release events
discovered by Hinode and can directly address the mechanisms of flux
emergence and cancellation.

The primary science goals of the high resolution Solar-C mission,
understanding the magnetic field throughout the solar atmosphere and
the release of energy through magnetic reconnection, wave dissipation,
and particle acceleration, lay the foundation for studying a wide range
of complex solar phenomena. Of particular interest are phenomena that
influence the near-Earth environment, such as coronal mass ejections
(CMEs), flares, and the solar wind.
Predicting the occurrence of CMEs and flares requires understanding
the storage and release of magnetic energy in the solar atmosphere. It
is clear that these occur on very different time scales.  Free
magnetic energy is built up slowly through the emergence, shearing,
and cancellation of flux in the photosphere. The release of energy, in
contrast, occurs rapidly, but what actually triggers an eruption has
not been firmly established.

The solar wind is another important component of space
weather. Coronal holes are the source of the fast solar wind (700 to
800~km~s$^{-1}$ at Earth orbit). The origin of the slow solar wind
(300 to 400 km s$^{-1}$) is unknown. Recently, instruments on Hinode
have detected persistent high speed outflows from large areas at the
periphery of many active regions
\citep[e.g.,][]{
2008ApJ...676L.147H}.  %Harra et al.%
These outflows are characterized by bulk shifts in the line profile of up to 
50~km~s$^{-1}$ and enhancements in the blue wing of up to 
200~km s$^{-1}$ \citep[e.g.,][]{2010ApJ...715.1012B}.  %Bryans et al.%
These outflows are of interest because they may lie on open field lines 
and contribute to the solar wind 
\cite{2003SoPh..212..165S}. %Schrijver & De Rosa%

Despite the extensive work on the outflows that has been done with
Hinode a number of important questions remain. Perhaps the most
important solar wind related question that Solar-C can address is how
these outflows are heated and accelerated. Temporally resolved
observations suggest that the outflows are composed of episodic events
that originate low in the solar atmosphere.  Solar-C will be better able to
study these outflows than other missions because of its wide
temperature coverage.  Solar-C can determine with precision at what
temperatures the outflows begin in the atmosphere and reveal, with its
high spatial resolution, the magnetic structuring at those
temperatures, providing the observations necessary to understand how
these outflows are formed and connect to the heliosphere as part of
the slow solar wind.

\section{LEMUR Scientific Goals and Requirements}
\label{sci-obj-req-lemur}
LEMUR provides diagnostic capabilities that are essential for
achieving the Solar-C science goals. It consists of a large VUV
telescope feeding a set of spectrometers and cameras capable of
observing simultaneously with high throughput all temperature ranges
(0.01 to about 20 MK) of the outer atmosphere with: (a) high spatial
sampling (about 200$\times$200~km$^2$, and down to 100$\times$100~km$^2$
in the 170 to 210 \AA\ band), (b) high spectral resolution 
(down to 2~km~s$^{-1}$ by line centroiding, 
ensuring dynamical scale coverage), (c) plasma 
diagnostic tools (such as electron density measurements), and (d) 
high time resolution (down to 10~s for raster
scans of a 4000~km wide region with 200~km sampling).

Moreover, LEMUR optics are designed to achieve the low scattering
performance necessary to observe, for example, the full life span of
transient events, and to explore faint reconnection outflow regions,
coronal holes, and far off-limb observations up to 0.5 solar radii
above the limb.  Finally, LEMUR is able to determine absolute coronal
velocities using wavelengths of chromospheric lines as zero-velocity 
references.

The solar chromosphere/transition region, with temperatures between
0.02~MK (upper chromosphere) and 1~MK (corona), strongly emits in
the VUV.  Strong lines from increasingly higher ionization stages
generally fall at shorter wavelengths.  The chromosphere and lower
transition region up to about 0.2~MK can be observed at wavelengths
$\geq$~900~\AA\ while the upper transition region between about 0.2~MK
and 1~MK and the corona $\geq$ 1~MK are best observed below
900~\AA. These relationships of temperatures to wavelengths allow us
to select wavelength ranges for LEMUR that maximize the range of
temperature sensitivity.  For the
first time, the full characterization of the entire transition region
and corona, and their linkages to the chromosphere, will be achieved
by a single instrument.  Moreover, a slit imaging assembly ensures that
the LEMUR slit spectra have excellent spatial co-alignment with the
photospheric and chromospheric magnetograms and filtergrams provided
by SUVIT at even higher resolution, ensuring that LEMUR observations
are always understood in the context of the surrounding lower
atmosphere.

VUV spectroscopy has several fundamental advantages over imaging, such
as provided by the AIA instrument aboard SDO.  
First, it can provide direct measurements of
electron temperatures, densities, and chemical abundances by using line
ratios, which, together with line profile/Doppler-shift measurements,
provide fundamental inputs to any theoretical modeling.  Second,
imaging instruments mostly show denser, cooling plasmas and provide very limited
information on the plasma that is being heated, and on plasma
velocities.  
Moreover, the information on temperature and plasma velocity
provided by spectroscopy improves the effective spatial resolution in two 
ways. One is the ability to see throughout the atmosphere at different
specific temperatures.
The solar corona is largely optically thin so multi-thermal structures 
can often not be resolved along the line-of-sight in broadband imaging data
due to the presence of additional spectral lines formed at different temperatures
just within a few Angstroms of the EUV spectrum
\citep{2006ESASP.617E..86D}. %Del Zanna et al.% 
The other is the ability to separate the emission in velocity space,
i.e., to measure plasma characteristics such as densities along the line-of-
sight for plasma flowing at different velocities.

Current and forthcoming VUV imaging spectrometers such 
as CDS, SUMER, and UVCS on SOHO, EIS on Hinode, and SPICE 
aboard Solar Orbiter either acquire simultaneously 
a few fixed restricted regions of the VUV spectrum or access a 
large spectral range by scanning in wavelength and therefore, 
unlike LEMUR, have limited instantaneous temperature coverage.
The required spectroscopy needs high sensitivity and
large telemetry volume, beyond what has been available with past and current
satellites, but  that will be provided by Solar-C.

\subsection{Impulsive energy release throughout the solar atmosphere 
{\it (Goals~1 and 3)}}
\label{sci-obj-req-lemur-goals-release}

Energy release is frequently highly dynamic, involving turbulent
motions and mass flows.  The magnitudes of the flows and turbulent
motions have complex temperature dependences, and the most dynamic
motions do not always occur at the highest temperatures.
The chromosphere and lower transition region are far more dynamic
than the quiet corona.  Waves, turbulent motions, and small eruptive
events are common in the chromosphere and transition region.

For example, possible heating via reconnection in the transition
region is suggested by observations of explosive events with plasma
flow velocities close to the Alfv\'{e}n speed (100~km~s$^{-1}$),
sometimes seen as bi-directional jets, in high resolution UV spectra
of transition region lines 
\citep{1997Natur.386..811I}. % Innes et al%%%%%%%%%%% 
These explosive events are in contrast to an apparent
steady average downflow motion in the transition region of $\approx$~2
to 25~km s$^{-1}$ and non-thermal motions of the order of 20~km
s$^{-1}$ \citep[e.g.,][]{1999A&A...349..636T}. %Teriaca et al.%%%%%%%

The signatures of energy transport and release from solar phenomena
never occur at just one temperature in the atmosphere.  For example,
in a solar flare radiation is emitted across the electromagnetic
spectrum from plasmas that span a temperature range from about 0.01~MK
to about 40~MK.  There is in addition a non-thermal component that
produces MeV particles sufficiently energetic to produce
nuclear reactions in the atmosphere.  As more quiescent examples,
magnetic loops in active regions span temperature ranges from about
0.01~MK to 3 -- 4~MK, and most spicules have temperatures
significantly less than 0.01~MK.  Every dynamical event exhibits
related multi-thermal phenomena.  This is true also for the diffuse
solar corona.  Thus, diagnosing multi-thermal plasmas is required to 
understand energy release in the solar atmosphere.

Waves are important throughout the solar atmosphere 
as a diagnostic tool for determining the conditions in the 
plasma they are propagating in.  
The rapidly-developing field of
chromospheric/coronal seismology aims to combine the
magnetohydrodynamic theory of waves in this structured
environment with emission line intensity, density, and
Doppler-shift measurements to determine detailed physical
parameters in the corona such as the magnetic field, the density
stratification, and heating and energy dissipation processes.
Moreover, since
waves and oscillations are often initiated by impulsive
heating processes, their observation provides an additional window into
how the corona is heated. Both kink-mode and longitudinal slow
mode standing and propagating waves have been detected in the
corona \citep[e.g.,][]{
1999ApJ...520..880A,     %Aschwanden et al.%  
2002ApJ...574L.101W},  %Wang et al.%
and EIS on Hinode has been able to detect small amplitude waves
\citep[e.g.,][]{2010ApJ...713..573M}. %Mariska, J.~T. & Muglach, K%
But, because of the
limitations of current instrumentation, their diagnostic
potential has not yet been fully realized. Doing this requires
spatially-resolved measurements at high time cadence in emission
lines formed over a range of temperatures along the basic
structural elements of the corona. The study of oscillation and
wave phenomena at all temperatures by LEMUR via seismological
techniques will allow us to determine poorly-known but crucial
 dissipation coefficients at sub-arcsec scales for the first time.

%%%%%%%%%%%%%%%%%%%%%%%%%%%%%%
\begin{figure}[t!]
\includegraphics[angle=270,width=120mm]{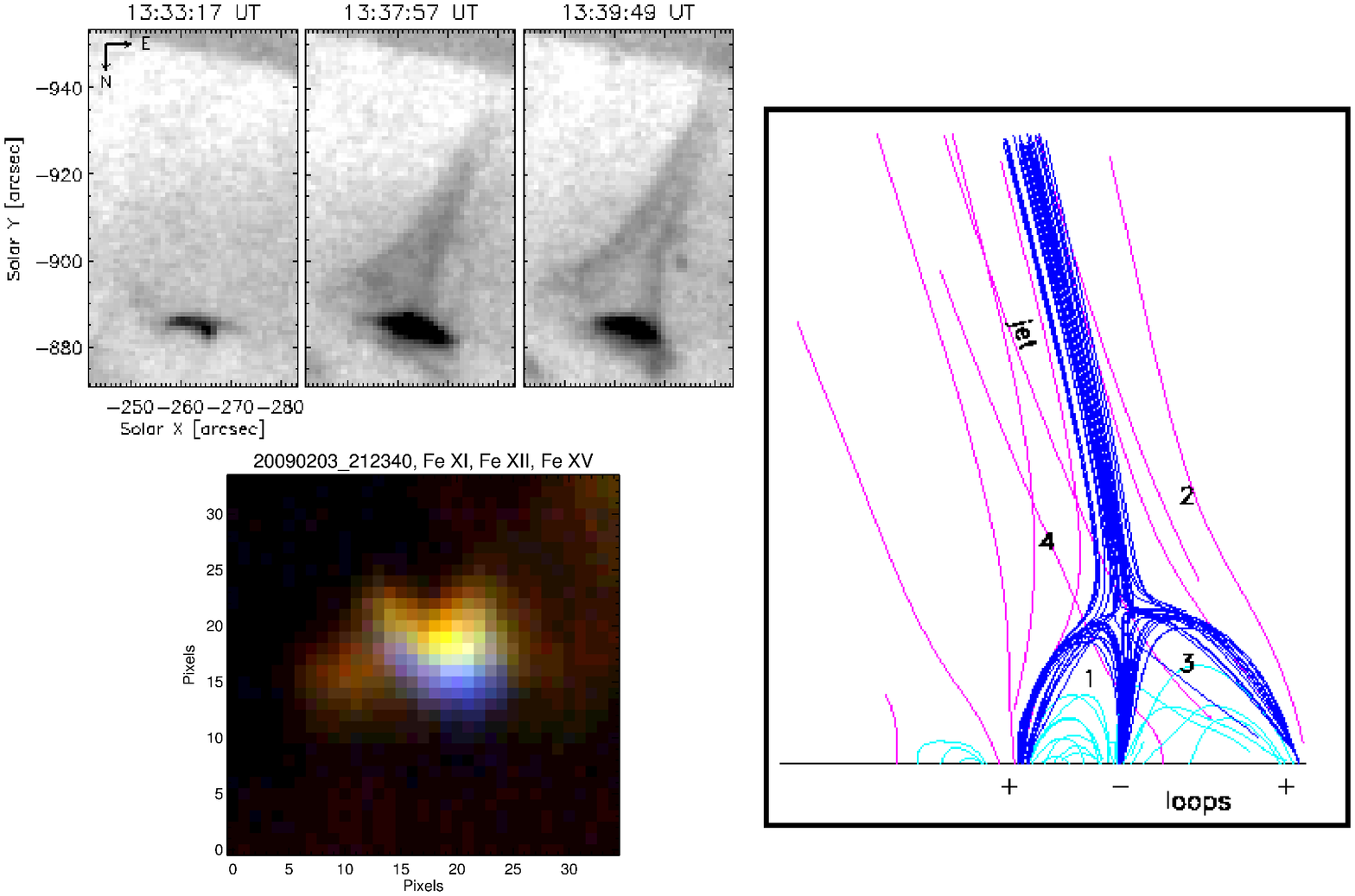}  %%PS
\vskip -1cm
\caption[]{\label{BP-jet} Top left: Hinode/XRT images of a jet from
  a bright point.  Lower left: Hinode/EIS composite image of a bright point. 
 The color code is red (Fe~{\sc xi}, 1.4~MK),
  green (Fe~{\sc xii}, 1.6~MK), and blue (Fe~{\sc xv}, 2~MK).  The bright point
  looks like a small loop, brightest at the top.  The separation of
  the two footpoints is only 5\,800~km.  The EIS pixel size is
  1\arcsec, so it is clear that the loop is not at all well-resolved.
  Right: A theoretical schematic of the magnetic configuration of a
  bright point region and a jet.}
\end{figure}
%%%%%%%%%%%%%%%%%%%%%%%%%%%%%%%%

Coronal hole bright points and their associated jets are an 
example of energy release over a small localized area. They appear to be
the result of magnetic reconnection when an erupting bipole encounters 
existing open magnetic flux. The simplicity of the bright points and 
their isolation from other heating events makes them attractive targets
for reconnection studies. Figure~\ref{BP-jet} shows XRT and EIS images 
of a jet and a bright point, respectively, and the schematic of a reconnection 
model. Hinode observations have shown that the heating and cooling typically 
occurred on timescales of tens of seconds, too short for the imaging
capabilities of EIS, but easily achievable with LEMUR.  
The spatial scales of jets should be resolved by LEMUR. 
The plasma conditions in the jets are crucial for testing reconnection models.

  LEMUR can
  determine the topology and morphology of the magnetically dominated
  structures ($\beta<1$) as well as the thermodynamic state of the
  plasma (its motions, density, temperature, turbulence, and
  abundances) that fills these structures.  This provides the required
  input for the computation of the large-scale coronal field by
  next-generation extrapolation techniques, which will include
  non-magnetic forces \citep{2006A&A...457.1053W} % Wiegelmann et al.%
  and plasma flows.

\subsection{Ubiquitous heating processes of the solar atmosphere {\it (Goal~2)}}
\label{sci-obj-req-lemur-goals-heating}

The source of the hot and tenuous plasma making up
the solar corona is still debated. The traditional view is that
heating occurs high in the corona, probably in an impulsive fashion by
reconnection events or by dissipation of steepened wave packets.  This
generates a downward thermal heat flux which heats and evaporates
chromospheric plasma into the corona.
Transverse magnetic waves can carry enormous amounts of energy upward
through the chromosphere \citep{
2007Sci...318.1574D,  %De Pontieu et al.%
2007Sci...318.1580C}. %Cirtain et al.% 
However, Solar-C's high resolution and spectroscopic capability at a wide range
of temperatures is required to trace the propagation, mode-coupling,
reflection, and dissipation of these waves as they cross the $\beta=1$ 
surface into the transition region and corona.  The ability to resolve their 
short periods ($<60$~s), small amplitudes ($<1\arcsec$), and
velocities  ($<20$~km~s$^{-1}$) is critical.

An alternative view is that a
subset of chromospheric spicules can heat up to coronal temperatures
as they rise and supply mass to the corona. These ``type II'' spicules
have typical widths of $\approx$ 0.3\arcsec, upflow speeds of 50 to
100~km~s$^{-1}$ \citep{2009ApJ...701L...1D}, %DePontieu et al.%%%%%
and generate observable features (e.g., asymmetries) in the line profiles
of several coronal spectral lines as observed at 2\arcsec spatial
resolution by EIS on Hinode. 
These findings have been recently substantiated by 
SDO/AIA observations
\citep{2011Sci...331...55D}.        %De Pontieu et al.%
However, spectroscopic observations at comparable spatial resolution
are necessary to establish whether mass is effectively supplied to the 
corona. 

\begin{figure*}[t!] 
%%% REMEMBER TO PROPERLY SET THE WIDTH IN THE FINAL VERSION.
\centerline{\includegraphics[angle=270,width=120mm]{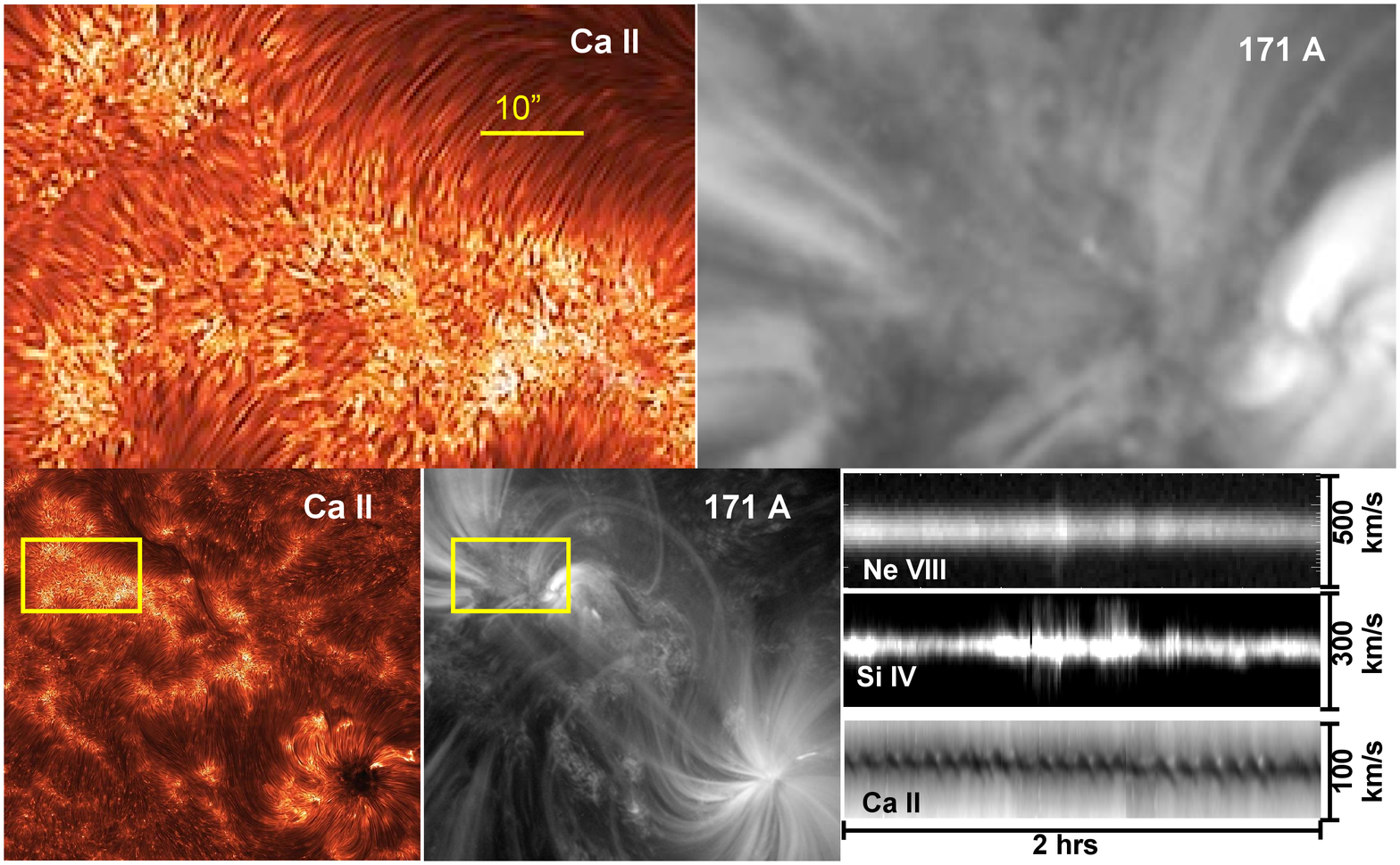}}  %%PS, PDF
\vskip -0.5cm 
\caption[]{\label{cauzzi} The chromosphere-corona connection. 
The top panels show cut-outs taken from 
the chromosphere (Ca~{\sc ii}, resolution similar to LEMUR) and 
the lower corona (171~\AA, $\sim$~1\arcsec\, resolution) images. 
The regions are outlined in yellow. The bottom right panel shows representative 
time series of chromospheric (Ca~{\sc ii}), transition region (Si~{\sc iv}), and 
lower coronal (Ne~{\sc viii}) line profiles. The chromospheric spectra have 
characteristic 3 to 5~min periods, the transition region brightenings occur in 
bursts of up to 30~min, while coronal variations are more diffuse.} 
\end{figure*}

The problems encountered in understanding the coupling
between the chromosphere and the corona are illustrated in
Figure~\ref{cauzzi}. The chromosphere has relatively large patches
(10\arcsec to 20\arcsec) containing many thin, bright spicules, and the
coronal structures are narrow, isolated loops that fade at their
footpoints.  There is little correspondence between the two.  A
comparable transition region image which can only be obtained by a
rastering spectrometer would take almost two hours to complete with
SUMER. Although the large-scale structures do not normally change on
this timescale, the small-scale dynamics are much faster.  LEMUR,
having more than 30 times higher throughput than SUMER in
transition region lines, will make a giant step forward.

In particular, LEMUR can connect in a systematic manner photospheric
phenomena with coronal phenomena because the chromosphere/transition
region interface will be observed with matching resolutions both below
(chromosphere and photosphere) and above (corona). The complementary
magnetic maps provided by the visible telescope on Solar-C assures
that the LEMUR observations can always be traced back to the plasma
and field conditions in the lower atmosphere.
%%%%%%%%%%%%%%%%%%%%%%%%%%% 
%%%%%%%%%%%%%%%%%%%%%%%%%%%%% 

SOHO/CDS observations 
\citep[e.g.,][]{2003A&A...406.1089D}  %Del Zanna & Mason%
have shown that active region emission between 2 and 3 MK is dominated 
by diffuse emission, i.e., unresolved at the current best (1\arcsec) 
resolution.
However, at lower temperatures, active regions contain `warm' loops
that reach temperatures no greater than about 1.4 MK and appear not
very far from being resolved.  This is seen in
1\arcsec\ EUV imaging (TRACE, SDO/AIA), and by measuring densities and
emission measures (hence path lengths and spectroscopic filling
factors) with EUV coronal spectrometers (CDS, EIS).  Indeed the
footpoints of these warm loops are anchored in magnetic field
concentrations of a few arcsec in size.  At higher temperatures,
in their core, active regions have hot (3 to 4 MK) loops,
frequently unresolved in imaging data.

Thus, current observations of the corona indicate the need for higher
spatial resolution data.  LEMUR's spatial resolution should
completely resolve at least the warm loops.  LEMUR can provide the
same high spatial resolution throughout the entire solar atmosphere
and provide by far the broadest diagnostics to measure the physical
conditions of loops.  

The fact that we resolve observationally a structure does not mean
that it could not be composed of unresolved magnetic threads.  Indeed
this is one of the assumptions about nanoflare heating of coronal
loops.  The plasma confined to these threads must be heated
sequentially in order to explain the longevity and densities of these
loops \citep{2010ApJ...713.1095W}.  %Warren et al.%%%%%%%
One issue is
whether plasma is heated in a steady or impulsive way. Detailed
measurements from Hinode/EIS are shedding some light into these
issues; however it still takes a very long time (tens of minutes) to
scan an active region loop from footpoint to footpoint.  The
much higher sensitivity and telemetry of LEMUR will allow us to
determine the relative roles of steady and impulsive heating in active
region structures and the validity of models such as the nanoflare
model.

\subsection{Solar flares {\it (Goal~3)}}
\label{sci-ob-req-lemur-goals-flares}

Solar flares form the high-energy tail of reconnection events. 
They are therefore ideal for studying how magnetic energy is dissipated via reconnection,
one of the important goals of the Solar-C mission. 

\begin{figure}[t!]
\includegraphics[angle=0,width=80mm]{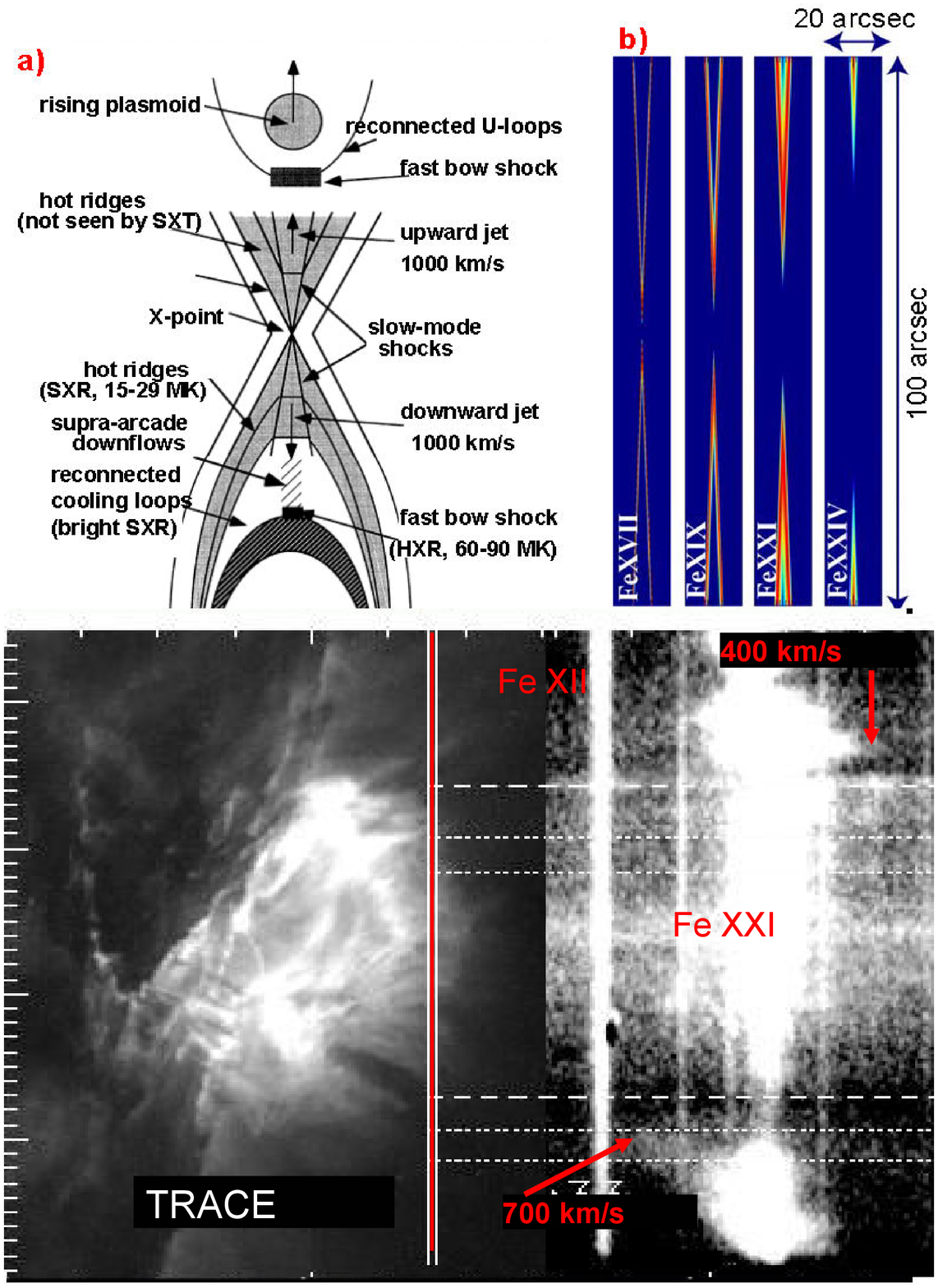} %%PS, PDF
\vskip -0.5cm
\caption[]{\label{imada_innes} The standard flare model.  a)
    cartoon showing the relationship between reconnection jets,
    shocks, and the flare arcade.  b) computation of non-equilibrium
    ionization structure in 1200 km~s$^{-1}$ jets.  c) large flare
    observed by TRACE and SUMER. The position of the SUMER slit is
    shown as a red vertical line on the TRACE image.}
\end{figure}

Figure~\ref{imada_innes}a presents a sketch of the standard flare scenario 
that successfully explains features seen in X-ray and EUV images, such as 
flare arcades, ribbons, plasmoid ejections, supra-arcade downflows, and 
hard X-ray loop-top emission. The predicted inflows, shocks and high-velocity 
jets, are central to this scenario but they must be faint because they 
have not been seen in images. If they exist, they should stand out as Doppler 
shifted features in LEMUR spectra. Also, rapid heating at the shocks might result 
in non-equilibrium ionization and different ion and electron temperatures 
which can be investigated with high cadence spectra. 
An example of the non-equilibrium ionization structure of slow-mode 
shocks is shown in Figure~\ref{imada_innes}b.  
The temperature evolution in the shocks is tracked by measuring the 
intensities of lines of highly ionized iron from Fe~{\sc xvii} through Fe~{\sc xxiv}.

The right part of Figure~\ref{imada_innes}c is an Fe~{\sc xxi} spectral 
image taken with SUMER, showing high velocity (400 to
1000~km~s$^{-1}$), high temperature (10~MK) plasma at many positions
along the spectrometer slit, above a flare arcade.  These observations
are highly suggestive of dynamics in or close to the reconnection
region, but SUMER was observing in sit and stare mode and the time
cadence was not better than 2 minutes (due to limited telemetry), so the
structure and dynamics could not be resolved.  Moreover, SUMER flare
observations were restricted to the off-limb because of its detector
characteristics.  Fast, high resolution Doppler mapping in the full
complement of hot Fe lines (Fe~{\sc xii} to {\sc xxiv}) within and
above the forming X-ray loops, together with high cadence X-ray or EUV
imaging observations, such as the context TRACE image shown in
Figure~\ref{imada_innes}, are required to determine the current sheet
geometry and to test the standard flare scenario.

More in general, it is
clear that magnetic reconnection in the corona occurs much faster than
the time scales predicted by the classic Sweet-Parker model, but what
drives this fast reconnection? Recent theoretical work has suggested
that it is the formation of magnetic islands in the current sheet that
leads to fast reconnection (e.g.,
\citep[][]{
2009PhPl...16k2102B,  %Bhattacharjee et al.%
2010ApJ...718...72E}). %Edmondson et al.%
To reconcile the
observations with theory and numerical models, the physical conditions
in and around the current sheet must be explored in detail. Of
particular importance are the inflow and outflow velocities. Since the
density in this region is close to that of the ambient corona and the
volume is small, such measurements are very challenging. This further
motivates the selection of multiple wavelength ranges so that the
brightest emission lines at critical temperatures can be observed
spectroscopically with high spatial resolution.

%%%%%%%%%%%%%%%%%%%%%%%%%%%%%%%%
%%%%%%%%%%%%%%%%%%%%%%%%%%%%%%%

\subsection{Large-scale dynamic phenomena {\it (Goal~4})}
\label{sci-obj-req-lemur-goals-large-scale}

\begin{figure}[b!]
\vskip -1cm
\includegraphics[angle=270,width=120mm]{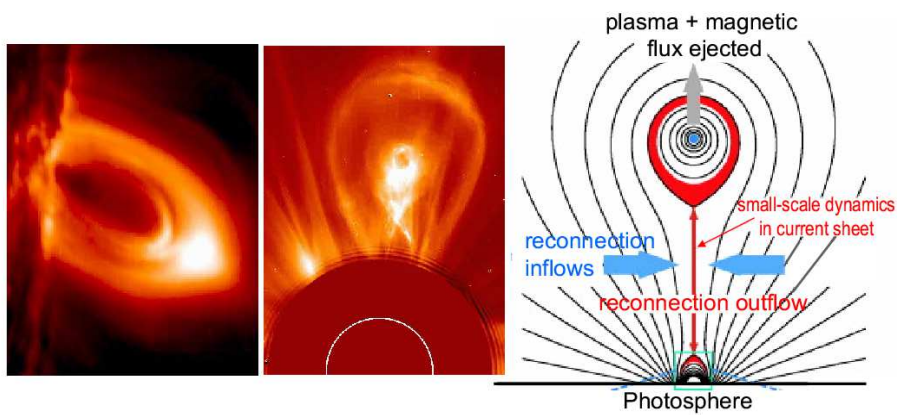} %%PS, PDF
\vskip -1cm
\caption[]{\label{CME-flare} Typical flare loops (EUV, left), coronal mass
ejection (CME, white-light, center) and schematic CME model (right).}
\end{figure}

CMEs and flares are closely related phenomena which begin near small sections 
of a photospheric polarity inversion line (Figures~\ref{imada_innes} and \ref{CME-flare}).  
Their onset is typically associated with the emergence of new
flux or with the dispersion and cancellation of old flux, generally with a high degree 
of fragmentation into the smallest observable scales.
Recent observational and computational studies indicate that a
topology change from a magnetic arcade to a flux rope occurs in either
case, before or in the early stages of the eruption \citep{
2007ApJ...666..532M, % Manchester 2007% 
2010arXiv1011.1227G}. % Green et al. 2010% 
Observing and quantifying this topology change is one of
the greatest challenges in space weather research. It requires
unraveling the signatures of reconnection at the smallest scales of
the flux fragments and learning how these organize into larger scales to form a
coherent magnetic structure of nearly active-region size that is able
to erupt. In the corona, such signatures have been seen as small-scale
brightenings and motions of brightened plasma along filaments and
filament channels, but a comprehensive observation of the evolution
from small to large scale has not yet been obtained. Moreover, on
theoretical grounds, the reconnection is expected to occur also at
photospheric and low chromospheric levels
\citep{1989ApJ...343..971V}. % van Ballegooijen & Martens (1989)% 
It may well be relevant also in the intermediate height range. The
capability to observe line shifts and broadenings across a wide range
of temperatures represents LEMUR's unique potential to discover
signatures of reconnection through several layers of the
atmosphere. The high spatial resolution of such measurements will
allow deeper insights into the topology changes preceding CMEs and
flares.
%%%%%%%%%%%%%%%%%%%%%%%%%%%%%%%
%%%%%%%%%%%%%%%%%%%%%%%%%%%%%%%

LEMUR's broad temperature coverage has the potential to discover links
between cool filament plasma trapped in the gradually destabilizing
flux and the ambient hot coronal structures (e.g., coronal
cavities). Since the reconnection that forms the erupting flux is
thought to often involve also the ambient flux, a linkage may
frequently occur, possibly only transiently, and hold important clues
to the genesis of the eruptions.

As soon as the eruptions are triggered, the unstable flux undergoes a 
rapid and huge expansion. Often the magnetic connections in the
source regions are further changed, transient coronal holes are
produced, and the corona is perturbed on a semi-global scale, 
which may involve the triggering of eruptions at remote locations, 
i.e., very complex space weather effects. Combining LEMUR and 
large-scale EUV (or X-ray) imaging observations allows following 
the evolution to the largest scales and mutually verifying and 
complementing the diagnostics,
for example by correlating line shifts and broadenings with images of
transient coronal holes. Only spectroscopic observations across a
substantial temperature range allow fully detecting the plasma
streaming into the legs of CMEs from the areas of transient coronal
holes, thus completing the diagnostics of the mass and energy budget
of the events 
\citep{2010SoPh..264..119A}. % Attrill et al. (2010)%

\subsection{The solar wind {\it (Goal~2)}}
\label{sci-obj-req-lemur-goals-wind}
    
\begin{figure}[t!]
\includegraphics[angle=270,width=120mm]{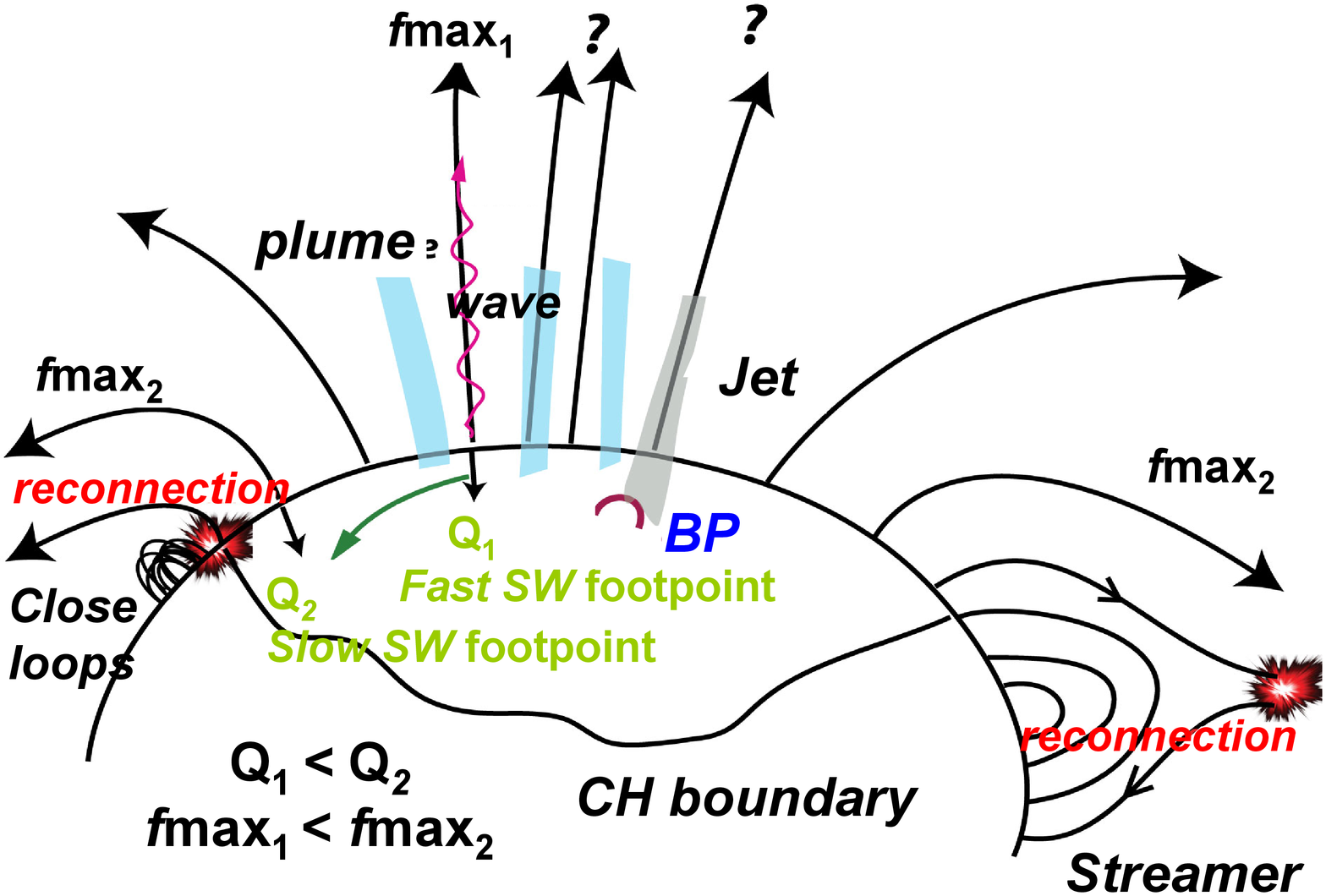} %%PS, PDF
\caption[]{\label{coronal-hole-schematic} A schematic showing
  various atmospheric structures found in coronal holes and possible
  sources of the fast and slow solar wind.  The quantities {\it f}max and Q
  refer to the rates of magnetic expansion and heating discussed for
  example by \cite{2009ApJ...691..760W} and
  references therein.}
\end{figure}

Despite many observations, it is unclear which structures and what
physical mechanisms are primarily responsible for the acceleration of 
the slow and fast solar wind. Figure~\ref{coronal-hole-schematic}
summarizes the state of our understanding.
LEMUR's capability of measuring the physical
parameters of the coronal plasma at greater heights and with higher
accuracy than previous missions, and in combination with the new
near-Sun in-situ measurements of Solar Orbiter and Solar Probe+, will
likely clarify the fundamental issues.  For example, LEMUR is capable
of measuring relative element abundances in different atmospheric
structures that can be related to in-situ measurements.  Element
abundances in the slow solar wind are different from those in the fast solar
wind, with the abundances of low-FIP (First Ionization Potential)
elements enhanced compared to high-FIP elements.  
The abundances of the fast solar wind are close to the photospheric 
abundances.  Thus,
abundance measurements can be used in conjunction with in-situ
measurements to locate the source regions  of the solar wind.

\begin{figure}[t!]
\includegraphics[angle=270,width=120mm]{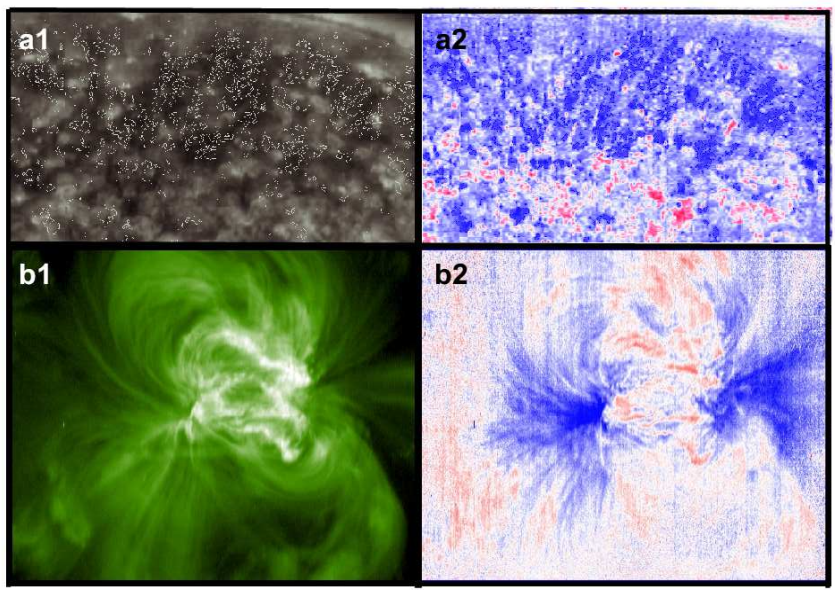} %%PS, PDF
\caption[]{\label{wind-flows} Wind outflows at the base of the corona: 
Panels {\it a1} and {\it a2}: A large portion of the north polar coronal hole imaged
in the Ne~{\sc viii} 770~\AA\ line with SUMER. Polar coronal holes are recognized 
as the sources of the fast solar wind.
Panels {\it b1} and {\it b2}: An active region imaged
 by EIS in the spectral line of Fe~{\sc xii} at 195.12 \AA. 
There is mounting evidence
that outflows from the edges of active regions contribute to the slow solar wind.
Panels {\it a1} and {\it b1} show integrated line radiances. Panels {\it a2} and 
{\it b2} show Doppler maps.  Blue indicates an outflow;
  red a downflow along the observer line-of-sight.  
Note that the extensive outflows occur in regions that are faint in Ne~{\sc viii}
in the polar coronal hole and in Fe~{\sc xii} in the active region.}
\end{figure}

It is well known from the early comparisons of solar X-ray flux and 
in-situ measurements of particle speeds that coronal holes (CH) are the
source of the fast solar wind
\citep[e.g.,][]{1973SoPh...29..505K}. %Krieger et al.%%%%%%%%%%
For polar
coronal holes, outflows of the order of 10 to 20~km~s$^{-1}$ have been
found at the boundaries of magnetic network cells from SOHO/SUMER
observations in the Ne~{\sc viii} 770~\AA\, line formed at
the base of the corona (T = 0.63~MK). The flows, as shown in
Figure~\ref{wind-flows}a, appear stronger in areas of reduced 
Ne~{\sc viii} emission 
\citep{2000A&A...353..749W}. %Wilhelm et al. 2000 %%%%% 
From SUMER measurements of C~{\sc iv} and Ne~{\sc viii} lines,
respectively, the wind starts flowing between 5~Mm and 20~Mm and flows
through magnetic funnels
\cite{2005Sci...308..519T}.    %Tu et al. 2005% 
Despite this progress, there is still a
substantial lack of knowledge as to which structures within coronal
holes and what physical mechanisms are primarily responsible for the
acceleration of the fast solar wind.  
Three candidates have been discussed as
the structures responsible: plumes, interplumes, and jets (see
Figure~\ref{coronal-hole-schematic}).

%%%%%%%%%%%%%%%%%%%%%%%%%%%%%
%%%%%%%%%%%%%%%%%%%%%%%%%%%%%

White light images of the solar minimum corona during eclipses reveal linear
structures, called plumes, rooted in coronal holes and extending
towards interplanetary space. These plumes could supply mass to the solar
wind; measurements from SOHO/CDS have suggested that they have near
photospheric abundances at their base
\citep{2003A&A...398..743D}. %Del Zanna et al.%%%%%%%%%
Observations in VUV
spectral lines formed at temperatures between 0.8 and 1~MK allow the
study of the plume roots where reconnection with newly emerging flux
has been suggested to cause their formation.
  
Plumes are known to be denser and cooler than the surrounding
interplume regions, however direct measurements with SUMER were
limited, mainly by sensitivity.  Spectral lines are observed to be
broader in interplume regions, hinting at preferential energy
deposition at these sites.  SUMER observed smaller outflow speeds at
the base of polar plumes, so interplume regions have been suspected to
be the source regions of the fast wind.  However, contradicting
results have been obtained from SUMER observations based on the
Doppler dimming technique applied to off-limb spectra \citep[e.g.,][]{
2003ApJ...588..566T, %Teriaca et al.% 
2003ApJ...589..623G}. %Gabriel et al.% 
LEMUR Doppler measurements with an accuracy of
$\leq$~2~km~s$^{-1}$ in coronal spectral lines are needed, together
with off-limb observations with the lowest possible level of stray
light (disk radiation scattered by telescope optics).  LEMUR can also
greatly extend and improve direct temperature measurements by
observing temperature sensitive line ratios (e.g., from Mg~{\sc ix})
as well as spectral lines separated considerably in wavelength, i.e.,
having substantially different excitation energies 
(e.g., from O~{\sc vi} and Fe~{\sc xii}). 
Moreover, comparing the chemical composition of
plume and inter-plume regions in the corona from LEMUR spectra with in
situ data by Solar Orbiter and Solar Probe+ at elevated latitudes could
establish which structures are the primary sources of the fast solar
wind.

Wave diagnostics are available for LEMUR as a tool for 
investigating the plasma properties of plumes and inter-plume regions.
Quasi-periodic fluctuations in intensity have been observed
in plumes and inter-plume regions, with periods of the order of 5 to
20~min 
\citep[e.g.,][]{2010arXiv1009.2980B}, %Banerjee et al., review%%%%%%%
and possibly are signatures of slow magnetoacoustic waves. 
However, accurate measurements of densities and temperatures
(as well as Doppler shifts) at much higher signal to noise ratio and
cadence (10 to 20~s timescales) than currently available in the
low corona, as can be provided by LEMUR, are needed to confirm this
scenario and to investigate higher frequencies.

The high spatial resolution of the Solar-C payload can provide a
substantial increase in our understanding of the polar regions, even
from 1 AU, on a path opened by the Hinode discoveries on the
polar magnetic landscape \citep{2008ApJ...688.1374T}.  %Tsuneta et al.
LEMUR, in particular, is able to
observe the poles with an effective (considering fore-shortening effects at
an 83$^\circ$ angle) resolution better than 2\arcsec\ (taking
advantage of the $\approx 7^{\circ}$ tilt of the solar rotational
axis).  The Solar-C location in the ecliptic plane makes on-disk
studies of the polar regions more difficult but it is ideal for
studies of the plasma conditions above the limb.  This makes the
mission highly complementary to Solar Orbiter, which will observe the
poles from about 34$^{\circ}$ above the ecliptic.  Solar Orbiter will
study the dynamics and evolution of on-disk polar regions (e.g., roots
of plumes) while Solar-C will look at the above limb effects of such
dynamics and evolution.

%%%%%%%%%%%%%%%%%%%%%%%%%%%%%
%%%%%%%%%%%%%%%%%%%%%%%%%%%%%%

Hinode observations have revealed persistent flows 
(see Figure~\ref{wind-flows}b) at the edges of active regions 
\citep[see,][]{2008ApJ...686.1362D}. %Doschek et al.%%%%%%% 
These coronal outflows are in part possibly related to open
field regions connected into the heliosphere 
\citep[e.g.,][]{2011A&A...526A.137D}  %%Del Zanna et al. %%
and could provide up to
25\% of the slow solar wind in some cases. They are therefore an
important feature to be studied in conjunction with close-by in-situ
observations from Solar Orbiter and Solar Probe+.  At present it is
impossible to connect these flows with transition region and
chromospheric structures, due to incomplete temperature coverage by
Hinode/EIS and its limited spatial resolution.  This can be achieved with
LEMUR.

\subsection{LEMUR Performance Requirements}
\label{sci-obj-req-req}
It is clear from the science discussion above that to understand
the phenomena to the depth that we desire, we need an imaging
spectrometer with much more sensitivity, temperature coverage,
and higher spatial and spectral resolution than is available from
all current instrumentation. One way to think about the design
requirements for LEMUR is to devise a set of basic observing
sequences that achieve the science goals we have outlined.
Table~\ref{scitrace} contains a list of desired observations
that address many of the LEMUR and Solar-C science goals along
with the kind of observing program that would acquire the data
necessary to achieve the science goal. These observations rely on
a small set of observing sequences listed at the bottom of the
table. Collectively, they define the design requirements for the
instrument.

%%%%%%%%%%%%%%%%%%%%%%%%%%
\begin{table*}
\caption[]{Science traceability matrix connecting LEMUR science
  goals to Solar-C goals.}
\label{scitrace}
\begin{tabular}{|p{0.2\textwidth}|p{0.5\textwidth}|c|c|c|c|}
\hline
& & \multicolumn{4}{c|}{Solar-C Goals}\\
\hline
LEMUR Science Goals & Sample LEMUR Observations & 1 & 2 & 3 & 4\\
\hline

Impulsive energy release throughout the solar atmosphere &

Onset, evolution and consequences of
small-scale magnetic flux emergence/cancellation.
Timing and flow geometry from chromosphere to corona of
e.g. explosive events, bright points, microflares.

LEMUR mode: QS Fast, QS Context &

\ding{52} & & \ding{52} & \\

& & & & & \\[-1.2ex]

& Waves in different structures. 
Determine wave mode, damping in e.g.
AR loops, prominences, plumes.

LEMUR mode: C Fast, SS Fast, AR Context, QS Context &

 & \ding{52} & \ding{52} & \ding{52}\\

\hline
Ubiquitous heating of the solar atmosphere&

Coupling of (quasi-)periodic processes in different atmospheric structures.
Determine temporal, spatial relationship between chromosphere and corona
in e.g. spicules, coronal loop foot-points in ARs.

LEMUR mode: QS Fast, AR Fast, Dynamics Fast &

\ding{52} & \ding{52} & \ding{52} & \\

\hline

Solar flares & 

Observe energy buildup in
highly-sheared areas of ARs. 

LEMUR mode: AR context &

\ding{52} & \ding{52} & \ding{52} & \ding{52}\\

& & & & & \\[-1.2ex]

& Observe energy release site structure, flows and evolution. Requires flare flag and position from XIT.

LEMUR mode: Flare, AR Fast &

  \ding{52} & \ding{52} & \ding{52} &\\

\hline

Large-scale dynamic phenomena &

Observe filaments and filament channels. 

LEMUR mode: AR Context, QS Context&

\ding{52} & \ding{52} & \ding{52} & \ding{52}\\

\hline

Solar wind &

Observe plumes and interplumes above the limb. 

LEMUR mode: QS Context, SS Fast, C Fast &

 & \ding{52} & \ding{52} & \ding{52}\\

& & & & & \\[-1.2ex]

&

Observe AR outflows.

LEMUR mode: AR Context, AR Fast &

 & \ding{52} & \ding{52} & \ding{52}\\

\hline
\end{tabular}

\begin{tabular}{|lccccc|}
 LEMUR mode & FOV & Slit width & Step Size & Cadence & T range \\

Quiet Sun (QS) Fast & $14\arcsec \times 100\arcsec$ & 0.28\arcsec & 0.28\arcsec & 50~s & 0.02 - 1.5~MK\\
QS Context & $100\arcsec \times 150\arcsec$ & 0.56\arcsec & 0.56\arcsec & 200~s & 0.02 - 1.5~MK\\
Corona (C) Fast & $150\arcsec \times 280\arcsec$ & 1\arcsec & 10\arcsec & 150~s & 0.02 - 20~MK\\
Sit and Stare (SS) Fast & $1\arcsec \times 280\arcsec$ & 1\arcsec & 0\arcsec & 3~s & 0.02 - 20~MK\\
Active Region (AR) Fast& $14\arcsec \times 150\arcsec$ & 0.28\arcsec & 0.28\arcsec & 25~s & 0.02 - 20~MK\\
AR Context& $280\arcsec \times 280\arcsec$ & 0.56\arcsec & 0.56\arcsec & 500~s & 0.02 - 5~MK\\
Flare& $28\arcsec \times 280\arcsec$ & 0.56\arcsec & 0.56\arcsec & 25~s & 0.02 - 20~MK\\
Dynamics Fast& $5.6\arcsec \times  100\arcsec$ & 0.28\arcsec & 0.28\arcsec & 10~s & 0.02 - 1.5~MK\\
\hline

\end{tabular}
\end{table*}
%%%%%%%%%%%%%%%%%%%%%%%%%%

Consideration of these desired observations, as well as many
others, and of the measured sizes, speeds, lifetimes, and
regions of propagation of the important solar phenomena using
available data has led to the design parameters of LEMUR listed 
in Table~\ref{table-requirements}. A
large primary mirror is needed to achieve high sensitivity and
spatial resolution. The high sensitivity enables high time
resolution. The mirror coatings are designed to provide high 
reflectivity over the full wavelength range covered by LEMUR. 
Image motion compensation is desirable due to the high spatial
resolution which requires high fidelity in tracking moving
features. The optical parameters are also chosen to maximize
spectral resolution in order to detect the small flows that have
been seen and that provide valuable insights into the observed
phenomena.

%%%%%%%%%%%%%%%%%%%%%%%%%%
\begin{table}[htbp]
\caption[]{\label{table-requirements}LEMUR instrumental requirements}
\begin{tabular}{ll}
\hline
Field & Required value \\ 
\hline & \mbox{}\\[-1.5ex]
Spatial resolution   & $\leq$0.28\arcsec \\
Spectral resolution & $\lambda/\Delta\lambda$ 17\,000 to 32\,000 \\ 
Doppler shift accuracy &  $\leq 2$ km~s$^{-1}$ \\
Doppler width accuracy &  $\leq 5$ km~s$^{-1}$\\
Temperature coverage & 0.01 to 20~MK \\
Field-of-view          & slit length 280\arcsec \\
raster coverage      & 300\arcsec\ (w/o re-pointing)\\
Exposure times      & $\leq$ 10~s (0.28\arcsec\ sampling) \\
                           & $\leq$ 1 s (1\arcsec\ sampling) \\
Mirror micro-roughness & about 3~\AA\ rms or better\\
\hline & \mbox{}\\[-4ex]
\end{tabular}
\end{table}

%%%%%%%%%%%%%%%%%%%%%%%%%%%%%%%%%%

The requirements summarized in Table~\ref{table-requirements}
can be met technologically by designing an
instrument with as few optical elements as possible.  The EIS
spectrometer, a two-optic system, provides a good baseline design for
LEMUR.  In order to achieve the resolutions and temporal requirements
stated above, the telescope mirror needs to be substantially
larger than the SUMER (rectangular, 9$\times$13~cm$^2$) and EIS 
(circular, 15~cm diameter) mirrors.
Further, a combination of a EUV multilayer (i.e., Mo/Si) and a coating 
such as B$_4$C is necessary to cover the large wavelength range needed 
to probe the required temperature range. For the same reason the instrument 
will not have a front filter. Thus, at shorter wavelengths, LEMUR will produce 
spectra and images completely free of the diffraction ghosts emanating from 
the mesh supporting the entrance filter.   
Low-scattering optics are necessary in order to explore faint
reconnection outflow regions, coronal holes, and to make 
off-limb observations at up to 1.5 solar radii. 
Accessing structures above the limb at the 
nominal spatial resolution of 0.28" (e.g., loop systems) requires a pointing 
mechanism moving the main mirror over a spherical surface (like in SUMER).
Precise co-alignment must be achieved with
other Solar-C on-board instruments.  The co-alignment must meet the
high spatial resolution of the spectrometer and requires a slit camera 
(slit imaging assembly) imaging the chromosphere at about 0.3\arcsec\ 
resolution.

\section{Mission profile: orbit and thermal considerations}
\label{sec-mission-profile}
The preferred orbit is an inclined geosynchronous 
orbit (GSO, altitude 36\,000~km, inclination 
$<$~30$^{\circ}$, period 1~day, similar to SDO), 
with a Sun-synchronous polar orbit (altitude 680 to 
800~km, inclination 97 to 98$^{\circ}$, period 
98~min, similar to Hinode) as backup. 
The GSO is preferred
because of the more stable thermal conditions and of 
the long periods of contact that can provide high volume 
telemetry and allow near real time scientific 
operations, which are required to meet science 
requirements.

The very favorable thermal environment provided by the GSO 
helps to meet LEMUR's requirements by implementing a passive, cold biased 
thermal system, optimized for insusceptibility to external 
heat sources. The adverse effect of Earthshine and albedo 
on LEMUR can be minimized by optimal radiator placement 
and orientation\footnote{In low Earth orbit, this effect would be much 
more severe and compensation heaters would be needed to dampen 
short-term temperature excursions.}.
In GSO, for several weeks 
daily eclipsing episodes of up to 70~min occur twice a year, 
when the S/C crosses the ecliptic. Compensation heaters will
minimize the post eclipse recovery time to less than 1~h.

An analysis based on a simplified 16-node simplified thermal 
mathematical model has shown that the thermal concept 
is uncritical, leaving room for adaptations during instrument 
design maturation. It should be noted that optical 
performance reasons do not allow conductive cooling, so 
that the mirror will run hot. 
The mirror of LEMUR will be figured and polished 
for an elevated constant operational temperature of 
$\sim$85\,$^{\circ}$C, a concept that optimizes figure 
stability and minimizes contamination. 
The feasibility of this concept is demonstrated by the SUMER primary 
mirror, successfully operating at $\sim$83\,$^{\circ}$C.

\section{LEMUR module design, performance characteristics, and key resources}
\label{sec-mod-payload}

The LEMUR design is driven by the instrument requirements given in 
Table~\ref{table-requirements} and meets them with ample margin.

%%%%%%%%%%%%%%%%%%%%%%%%%%%%%%%%%%%%%%%%%%%%%%%%%%%%%%%%%%%%%%%%%%%%%%%%%%%%	
\subsection{LEMUR module description}

\begin{figure*}[t]
\vskip -0.30cm
%%% REMEMBER TO FIX THE WIDTH FOR THE FINAL VERSION
\centerline{\includegraphics[angle=270,width=120mm]{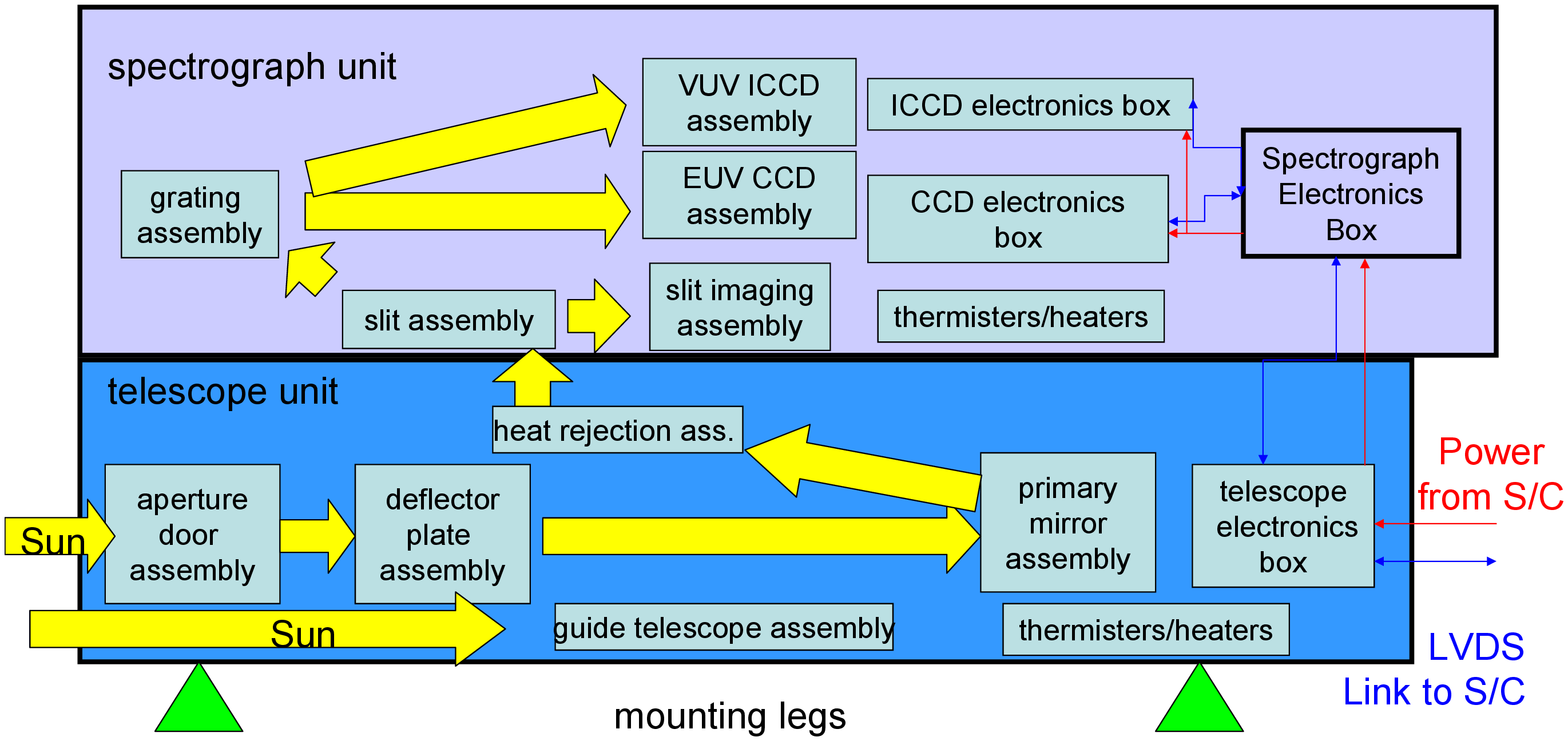}} %%PS, PDF
\vskip -2.0cm
\caption[]{\label{fig:mp1} LEMUR physical block diagram showing major LEMUR 
components and digital communication links. The telescope electronics box, 
provides power and digital interfaces (via low-voltage differential signaling, 
LVDS links) to the S/C on one side and to the spectrograph electronics box
on the other.}
\end{figure*}

The LEMUR module consists of a telescope unit and a spectrograph unit. 
A physical block diagram of the LEMUR module is given in Figure~\ref{fig:mp1}. 

\begin{figure*}[t!]
\vskip -0.3cm
%%%REMEMBER TO FIX THE WIDTH FOR THE FINAL VERSION
\centerline{\includegraphics[angle=270,width=120mm]{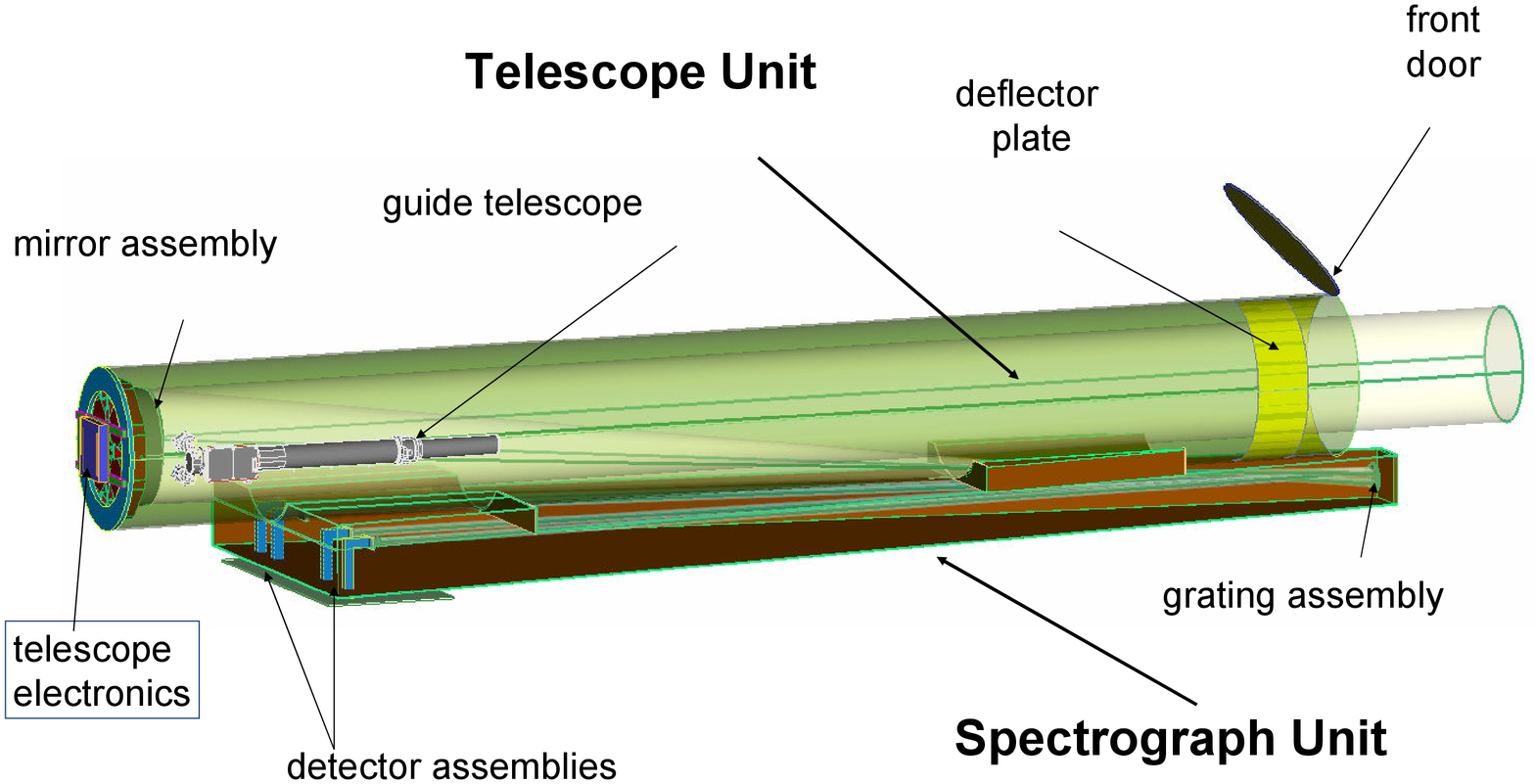}} %%PS, PDF
\vskip -1.5cm
\caption[]{\label{fig:mp2}  The opto-mechanical layout of LEMUR. Incoming
solar light enters the instrument from the right, through the front door. }
\end{figure*}

The module will be mounted with isostatic mounting legs to a mechanical 
interface provided by the Solar-C spacecraft.  The mounting will be designed to
provide thermal independence from the spacecraft. 
The Telescope unit structure is the main support structure of  the LEMUR module 
and has provisions to mount the Spectrograph Structure.
The opto-mechanical layout of the LEMUR system is shown in Figure~\ref{fig:mp2}. 
The telescope unit consists of a front door assembly, the single-mirror 
telescope assembly (an off-axis parabola with pointing capabilities), the guide 
telescope, the telescope electronics, the heat rejection assembly and the 
deflector plate assembly.
  
The spectrograph unit houses the slit assembly, the slit 
imaging assembly, the grating assembly, the focal plane assemblies, CCD electronics 
box, Intensified CCD (ICCD) electronics box and detector radiator assembly.  The 
Telescope unit and the Spectrograph unit are designed to be assembled and tested 
separately. They can be mated with each other for final testing and calibration 
before integration to the spacecraft (S/C).\\

%%%%%%%%%%%%%%%%%%%%%%%%%%%%% 
%%%%%%%%%%%%%%%%%%%%%%%%%%%%%

\noindent {\bf LEMUR telescope unit} -- The telescope unit provides the 
solar image in its focal plane, the entrance to the 
spectrograph unit, and provides image stabilization and pointing capabilities. The 
mirror assembly consists of the telescope mirror and a combined pointing and raster 
scanning mechanism. It is used to compensate for residual image motion and to provide 
coarse pointing of the telescope, as well as stepwise scanning to produce raster 
images. The mirror produces an image in the center of the focal plane of the 
parabola, to feed the spectrograph unit. Before the slit plane, a pre-slit heat 
rejection mirror collects more than 95\%\ of the heat of the solar image to be 
directed to a heat dump, whereas only a small fraction of the image will 
pass towards the slit.
The open telescope design requires a front door to protect the instrument during ground 
and launch operations. Electrostatic deflection plates, close behind the front door 
in the telescope baffle, protect the mirror coating from incoming solar wind 
particles. A guide telescope provides a signal for image motion stabilization.\\

\noindent {\bf LEMUR  spectrograph unit --}
The spectrograph unit accepts the light from the telescope passing through the entrance 
slit towards the grating and the detector assemblies. The slit assembly consists of slits of five 
widths mounted on a translation stage mechanism used to select the 
slit for a given observing program. The slit assembly also contains a focus 
adjustment capability. The grating assembly consists of two gratings, one for the 170 to 
210~\AA\ EUV band (SW) and the other for three bands at wavelengths above 
480~\AA\ (LW-1: 695 to 815~\AA, LW-2: 965 to 1085~\AA, LW-3: 1150 to 1270~\AA). 
LW-2, and LW-3 are expected to record some particularly intense second order lines. 
The grating assembly is mounted 
on a focusing mechanism. Each grating is ruled, figured and coated to optimize the 
optical performance of its respective channel. The dispersed slit images are projected 
onto the focal plane cameras, the SW camera with a thin foil filter and the three 
solar-blind intensified cameras, LW-1, LW-2, and LW-3.

Table~\ref{table:mp1} summarizes unit and assembly key characteristics.  Description and 
design characteristics of the LEMUR optics, detectors, structure and electronics are 
provided in the sections below.

%%%%%%%%%%%%%%%%%%%%%%%%%%%%% 
% FIG OPTO-MECH LAYOUT
%%%%%%%%%%%%%%%%%%%%%%%%%%%%% 
%%%%%%%%%%%%%%%%%%%%%%%%%%%%%
\subsection{LEMUR detailed design}

The LEMUR design follows that pioneered by the SUMER/SOHO and EIS/Hinode 
instruments. The telescope unit is contained in a precision manufactured, low 
expansion, light weight CFRP structure. Behind the entrance door is
a long optical baffle completed by an aperture stop with optical power (reflective 
and curved) for stray light reduction and heat management. Inside the 
telescope baffle will be a deflector plate assembly, a plate arrangement to protect 
the primary mirror from ambient charged particles. A static voltage is applied across the 
two plates such that solar wind particles will be deflected from their path 
towards the mirror.
The mirror will be mounted to a mirror assembly with radiative coupling to a radiator 
for cooling. The mirror assembly provides mechanisms for coarse pointing and fine 
rastering movements. The coarse pointing provides movement on a spherical surface around 
the focal center in north-south and east-west directions with a range of 3000\arcsec. 
The fine pointing capability allows image motion stabilization and raster scanning 
in a 300\arcsec range.
The telescope electronics box provides control of the telescope mechanisms and 
sensors and handles the signal of the guide telescope. It provides the digital and 
power interfaces to the S/C on one side and to the spectrograph electronic box 
on the other. During operation of the LEMUR module, it is controlled by the 
spectrograph electronic box.
The off-axis paraboloid telescope projects the solar image onto a slit assembly. 
Before the slit is a heat rejection assembly, a curved mirror to reject all 
but a 250\arcsec$\times$400\arcsec patch of the solar image. The rejected light is
transferred to a heat dump or reflected out to space. 
The heat rejection assembly belongs to the 
telescope unit while the slit assembly is part of the spectrograph unit. The 
spectrograph structure is a precision manufactured CFRP honeycomb panel structure, 
to be mounted on the telescope unit.

The slit assembly contains a double mechanism to select one of several slits on 
a slit substrate and a linear translation to move the slit into best focus of the 
telescope. Several slits with different widths will be available (e.g., 
0.14\arcsec, 0.28\arcsec, 0.56\arcsec, 1\arcsec and 5\arcsec, the latter to be used 
mostly for calibration purposes).

The slit jaws reflect the solar image passing through the pre-slit towards the 
slit imaging assembly: a small monochromatic camera with 
relay optics to provide a context image of the slit position in a 
200\arcsec$\times$300\arcsec field of view and a resolution of 0.3\arcsec. The wavelength 
band will be selected to fulfill co-registration requirements with the other SOLAR-C 
instruments.

The spectrograph splits the beam into the SW and LW channels by the two 
gratings which are tilted by a small angle oppositely with respect to the optical 
dispersion plane. Thus, each channel incorporates a highly corrected, single-grating 
imaging spectrograph.
First and second order spectral lines are chosen by a special selective 
photocathode coating on the LW detectors (see below). The gratings are housed 
on a linear translation mechanism with 2~cm range for focus adjustment, whenever 
necessary, to provide dispersed images of the entrance slit in the detector plane. The two 
focal plane assemblies are imaging arrays, highly efficient for each spectral 
passband. The SW detector is constructed with two butted, back-side thinned CCDs 
(2k $\times$ 2k format) and a thin metal foil filter to block visible radiation. The LW 
focal plane arrays are three CCD units coupled with a microchannel plate (MCP) 
intensifier.
The spectrograph electronic box (E-box) drives all mechanisms and 
cameras of the spectrograph units and provides 
commands to the telescope E-box. 
All design characteristics of the LEMUR module are summarized in
Table~\ref{table:mp1}.

%%%%%%%%%%%%%%%%%%%%%%%%%%%%%%
% TABLE: LEMUR module design characteristics
%%%%%%%%%%%%%%%%%%%%%%%%%%%%%%

\vspace{2ex}
\begin{table*}
\begin{center}
\caption[]{\label{table:mp1} LEMUR module design characteristics}
\begin{tabular}{ll}
\hline
Unit/assembly             & Nominal value/description \\

\hline & \mbox{}\\[-1.5ex]
\multicolumn{2}{c}{Telescope unit}\\[0.5ex]

Telescope type            & Single-mirror off-axis paraboloid \\

Optical characteristics   & 30~cm diameter, 360~cm focal length, $\lambda$/75 rms figure, \\
                                 & $<$3~\AA\ rms micro-roughness, broadband VUV coating \\
                                 & low-expansion substrate (Zerodur or ULE) \\

Deflector plate  & Two plates with static HV supply \\

Baffles/aperture stop   & Built as heat rejection mirror\\

Heat rejection pre-slit  & Curved mirror in front of the slit, 250$\arcsec\times400\arcsec$ central aperture\\

Telescope  pointing          & Coarse pointing capability: 3000\arcsec range in pitch and yaw \\
                                     & $<1 \arcsec$ steps with $10\arcsec$ accuracy. \\
                                       & Fast raster/fine pointing mechanism: range $300~\arcsec$ \\
                                       &  image motion in pitch and yaw, $0.028\arcsec$ digitization, \\
                                       & $>$10 Hz response\\

Guide telescope               & $<0.05 \arcsec$ accuracy at 10 Hz, range $>2000\arcsec$ \\

Telescope E-box             & Control of telescope mechanisms and sensors \\
                                    & Digital and power interfaces to the S/C \\

Front door                     & Reclosable front door \\
Telescope structure         & Precision manufactured CFRP structure \\
\hline & \mbox{}\\[-1.5ex]
\multicolumn{2}{c}{Spectrograph unit}\\[0.5ex]
Spectrograph design     & Ellipsoidal variable line space (EVLS) gratings \\
Slit assembly                      & Linear translation, 5 slits on a slit plate, 0.14$\arcsec$, 0.28$\arcsec$, 0.56$\arcsec$, 1$\arcsec$, 5$\arcsec$ \\
                                         & $>$2~cm focus adjustment along telescope chief ray \\
                                         & Slit jaws reflective to feed the slit imaging assembly \\
Grating assembly                & Houses the  SW and LW gratings \\
                                         & $>$2 cm focus adjustment along telescope chief ray \\
CCD focal plane assembly    & SW CCD with radiative cooling, 2 butted 2048$\times$2048 \\
                                        & 13.5~$\mu$m pixel with shutter mechanism, thin metal foil filter \\
ICCD focal plane assembly  & 3072$\times$2048, 20~$\mu$m pixel LW ICCD \\
                                      & bare and CsI or KBr coated MCP \\
                                      & protective vacuum door (optional)\\
CCD E-box                      & Controls and reads the CCD detectors \\
ICCD E-box                       & Controls and reads the ICCD detectors, \\
                                       & provides MCP HV  \\
Slit imaging assembly         & Relay optics with   $>$200$\arcsec\times300\arcsec$ FOV \\
                                    & $<0.3\arcsec$/pixel plate scale, CCD detector \\
Spectrograph structure       & Precision manufactured CFRP honeycomb panel structure \\
Spectrograph E-box & TM and TC handling, data processing, camera operations\\
\hline  & \mbox{}\\[-1.5ex]
\end{tabular}
\end{center}
\end{table*}

%%%%%%%%%%%%%%%%%%%%%%%%%

\subsubsection{LEMUR optical design}

\begin{figure}[t!]
\vskip -0.6cm
\includegraphics[angle=0,width=120mm]{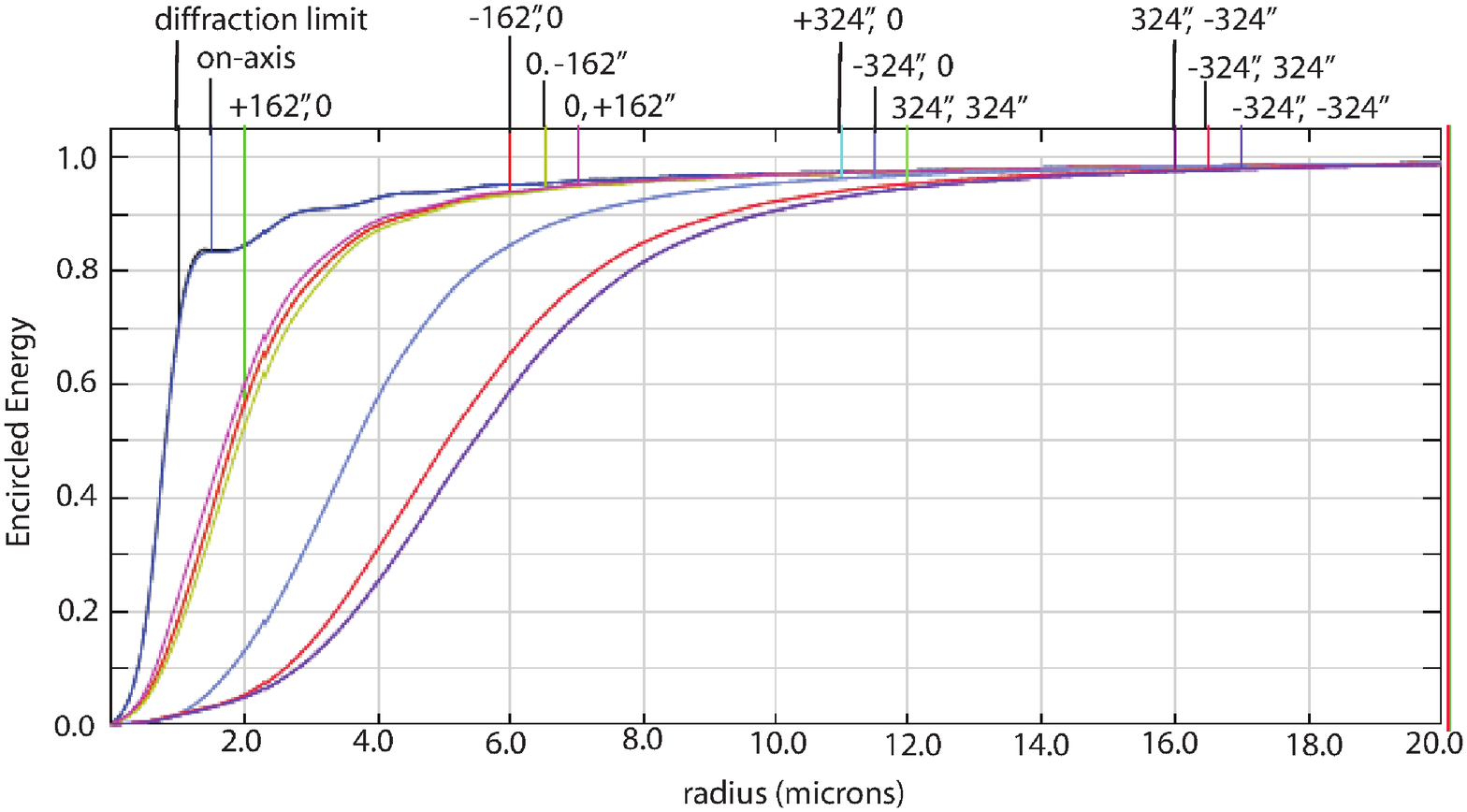} %%PS, PDF
\vskip -1.2cm
\caption[]{\label{fig:rayt} Imaging performance of the LEMUR telescope.  
Encircled energy radius in microns  are given for the off-axis paraboloid 
mirror at 1216~\AA\ for on-axis, $\pm$160, and $\pm$320 arcsec field angles.}
\end{figure}

The LEMUR optical design follows the two element design pioneered by the EIS
instrument aboard Hinode. 
The telescope mirror is a 30~cm diameter section of a parabola, 
placed 22.5~cm off axis, with 360~cm focal length. The mirror produces a plate 
scale of 17.5~$\mu$m/\arcsec at the location of the spectrograph entrance slit. 
The imaging performance of the telescope is shown in Figure~\ref{fig:rayt}.  
The optical characteristics of the telescope are given in Table~\ref{table:mp1}.  

%%%%%%%%%%%%%%%%%%%%%%%%%%%%% 
% Telescope performances
%%%%%%%%%%%%%%%%%%%%%%%%%%%%% 
%%%%%%%%%%%%%%%%%%%%%%%%%%%%%

\begin{figure*}[t]
\vskip -2.cm
%%%REMEMBER TO FIX THE WIDTH FOR THE FINAL VERSION
\centerline{\includegraphics[angle=270,width=120mm]{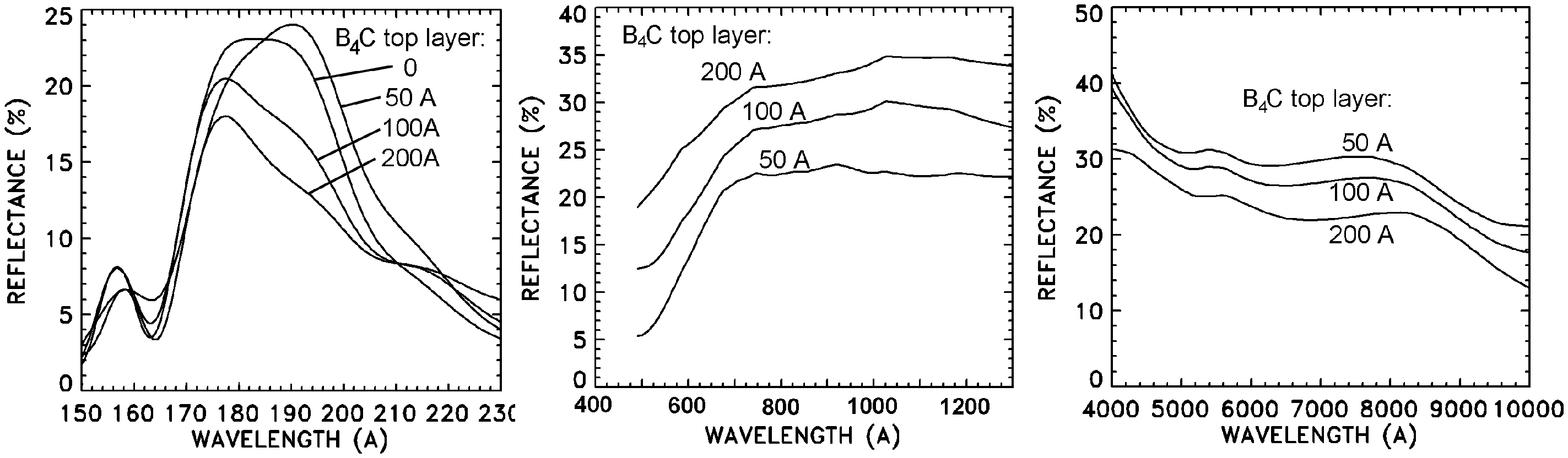}} %%PS, PDF
\vskip -2.4cm
\caption[]{\label{fig:mlt} LEMUR broadband Mo/Si multilayer mirror coating with B$_{4}$C
top layer with thicknesses of 50~\AA, 100~\AA, and 200~\AA. The reflectance is shown for 
the SW band (left), LW band (middle), and the Vis-NIR range (right).}
\end{figure*}

The LEMUR mirror coating is designed to have high normal-incidence reflectance 
in the VUV wavelength ranges 170--210 \AA\ and $>$480 \AA. The LW channel 
($>$480 \AA) waveband is reflected from a top B$_4$C layer. The SW channel 
wavelengths penetrate the B$_4$C layer and the 170--210 \AA\ wavelengths are 
reflected by a multilayer interference coating consisting of alternating 
layers of Si and Mo, traditional materials with extensive flight heritage.  
A unique reflective coating combination will be chosen for the entire 
area of the mirror as different combinations of coatings over different 
areas of the mirror might stress it and degrade its optical quality.
Shown in Figure~\ref{fig:mlt} is the reflectance calculated for three assumed 
B$_4$C top layer thicknesses (50~\AA, 100~\AA, and 200~\AA). The trade-off is 
that increasing the B$_4$C  thickness has the beneficial effect of increasing 
the $>$480~\AA\ reflectance but reduces the reflectance at shorter wavelength. 
The visible and near-IR reflectance is also shown since it affects the 
temperature of the mirror that accepts the solar heat load and also affects 
possible scattered out-of-band light.
The mirror is required to have micro-roughness of 3~\AA\, rms or 
better\footnote{Mirrors with $<$~2~\AA\, rms micro-roughness are routinely 
produced by companies such as Zeiss for EUV lithography. Moreover, at longer 
wavelengths, samples with $<$~2~\AA\, rms micro-roughness have been 
produced for the SPICE instrument to be flown on Solar Orbiter.}, 
significantly improving (of a factor $\approx$~4) stray light performances with 
respect to SUMER (6~\AA\, rms, mid-frequency error). 

%%%%%%%%%%%%%%%%%%%%%%%%%%%%% 
% Broadband reflectance
%%%%%%%%%%%%%%%%%%%%%%%%%%%%% 
%%%%%%%%%%%%%%%%%%%%%%%%%%%%%

The slit selects a portion of the solar image and passes it onto the SW and LW 
gratings.  Similar to CDS/SOHO, these two grating halves are mounted in close 
proximity. The gratings are figured and coated to optimize the image quality 
and efficiency of each passband. The 
highly corrected off-axis elliptical gratings with variable line spacing give 
excellent image quality at the focal plane across all wavelength ranges.  
Table~\ref{table:mp3} gives the prescriptions of the optics and the 
spectrograph characteristics. 

%%%%%%%%%%%%%%%%%%%%%%%
% Table spectrograph optical prescriptions
%%%%%%%%%%%%%%%%%%%%%%
\vspace{2ex}
\begin{table*}
\begin{center}
\caption[]{\label{table:mp3}
LEMUR spectrograph optical prescription}
\begin{tabular}{lll}
\hline
& \multicolumn{1}{c}{SW waveband} & \multicolumn{1}{c}{LW wavebands} \\
\hline & \mbox{}\\[-1.5ex]

Spectral ranges & 170--210 \AA & 1$^{\rm st}$ order:  695-815 \AA, 965-1085 \AA, 1150-1270 \AA \\
               & & 2$^{\rm nd}$ order:  482-542 \AA, 575-635 \AA \\

Dispersion & 4200 grooves/mm &  1200 grooves/mm \\
                      & 280 cm grating to detector  & 414.8 cm grating to detector  \\
                             &~10 m\AA/pixel &~40 m\AA/pixel\\
Plate scale & 0.14 \arcsec/13.5 $\mu$m pixel & 0.14\arcsec/20 $\mu$m pixel\\

Slit length &	280\arcsec (2000 pixels) & 280\arcsec (2000 pixels) \\
Magnification & 5.5 & 8.1 \\
\hline & \mbox{}\\[-1.5ex]
\end{tabular}
\end{center}
\end{table*}
%%%%%%%%%%%%%%%%%%%%%%%%%
The coating for each grating half is optimized for its particular waveband.  
The LW grating half is coated with 200\AA\ of B$_4$C.  The SW grating half 
is coated with Mo/Si bilayers without the B$_4$C top layer.

\begin{figure}[t]
\includegraphics[angle=0,width=120mm, trim = 0mm 50mm 10mm 50mm, clip]{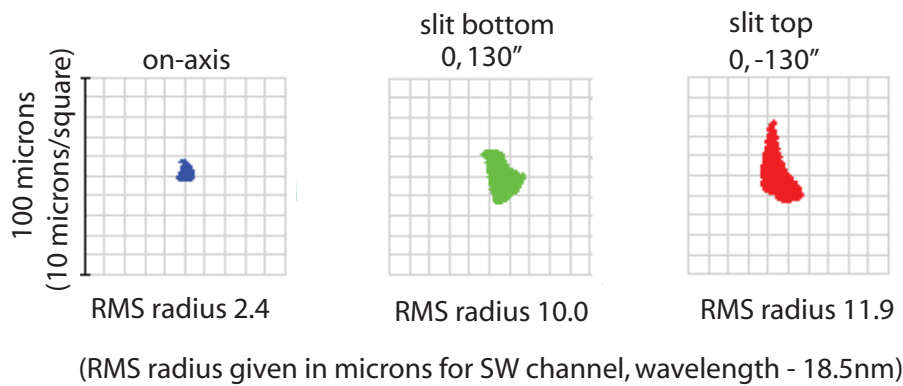} %%PS
\caption[]{\label{spot-sw} Combined telescope and SW spectrograph imaging 
performance at 185 \AA\ wavelength.  Spots shown are for slit center, 
slit top and slit bottom position.}
\end{figure}

\begin{figure}[t]
\includegraphics[angle=0,width=120mm, trim = 0mm 20mm 0mm 10mm, clip]{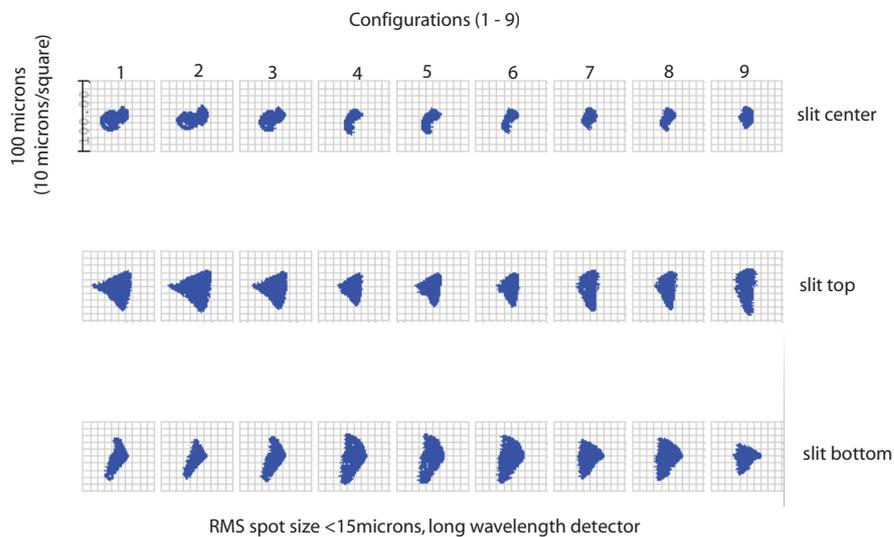} %%PS
\caption[]{\label{spot-lw} Combined telescope and LW spectrograph imaging 
performance.  Configurations 1 to 3 are for LW-1; 
configurations 4 to 6 are for LW-2; and configurations 
7 to 9 are for LW-3.  The top row of spots are at the center 
of the slit.  The second and third rows are the spots at the end of the 
slit. The scale of the square field is 100 $\mu$m on the side.  Focal plane 
pixel size is 20 $\mu$m.}
\end{figure}

The combined telescope and SW spectrograph performance at the central 
wavelength of 185 \AA\ at the slit center and extreme ends of the slit is shown in 
Figure~\ref{spot-sw}.  
The combined telescope and LW spectrograph performance is shown 
in Figure~\ref{spot-lw} for all wavelength ranges.
The LEMUR performance improvements are the result of a remarkable 
extension in spectrograph design which incorporates an off-axis Ellipsoidal 
Variable Line Space (EVLS) grating to 
produce the necessary resolution on the critical spectral, spatial, and temporal 
scales. The key to the success of this ellipsoidal grating concept is to orient 
the ellipse so that its two foci lie at the slit and at the detector.  
The segment of the ellipse actually used by the grating is off its normal axis. 
The remaining parameters of the EVLS grating are optimized using ZEMAX.  
The high grating magnification and the two elements optical design allow
LEMUR to reduce its size relative to a conventional spectrometer with similar 
spatial resolution by a factor between 3 and 5. 
EVLS gratings do not require new technology development.

%%%%%%%%%%%%%%%%%%%%%%%%%%%%% 
% FIGURE SW RAY-TRACE
%%%%%%%%%%%%%%%%%%%%%%%%%%%%% 
%%%%%%%%%%%%%%%%%%%%%%%%%%%%%

%%%%%%%%%%%%%%%%%%%%%%%%%%%%% 
% FIGURE LW RAY-TRACE
%%%%%%%%%%%%%%%%%%%%%%%%%%%%% 
%%%%%%%%%%%%%%%%%%%%%%%%%%%%%

\subsubsection{Focal plane assemblies}

The baseline detector is a CCD from the E2V 42 family which has extensive 
flight heritage.   During the conceptual design phase, the LEMUR team will 
study various detectors including the radiation hardened APS detectors and 
electronics presently under development for the Solar Orbiter mission.\\ 

\noindent {\bf CCD focal plane assembly -- } Similar to Hinode/EIS, it 
consists of two backside thinned CCD detectors.  These detectors are 
closely butted together for a minimal gap.  The focal plane incorporates 
a 1000 \AA\ thick aluminum foil filter to reject the visible radiation.  
This camera assembly is equipped with a focal plane shutter. The CCDs 
are thermally isolated and strapped to an external radiator to be 
cooled to $-$60\,$^{\circ}$C.\\

\noindent {\bf ICCD focal plane assembly -- } The LW detectors 
consist of three independently operated intensified CCDs. 
ICCD operation is similar to previous ICCDs flown on SOHO/CDS and the SERTS 
instrument. The MCP intensifier provides high efficiency and solar 
blindness such that no focal plane filters are needed.  Coupling 
between intensifier and CCD is made by fiber optic taper from a 20 $\mu$m
MCP pixel to a 13.5 $\mu$m CCD pixel (1.48:1).
The MCP will be coated with photocathode material of cesium iodide 
(CsI) or potassium bromide (KBr) in selective locations of the spectral ranges to 
enhance sensitivity of first-order lines against second-order lines that are, this 
way, largely suppressed. Five spectral ranges over LW-3 will be 
left uncoated to allow four selected 2$^{\rm nd}$ order lines of particular relevance to 
dominate over the first order lines. A fifth area of LW-3 will be left uncoated to 
reduce the extremely strong signal from the hydrogen Lyman $\alpha$ line. 
The MCP intensifier may be protected until scientific operations start by a 
protective door that can be opened in orbit and closed on demand.
The CCDs will provide an array of 3072 $\times$ 2048 pixels 
with 20~$\mu$m size.
The exposures are electronically controlled by switching the MCP voltage 
and the CCD can be operated without mechanical shutter. The CCDs can be 
passively cooled to $-$60\,$^{\circ}$C. 
This requirement will be revisited during the initial design phase.

\subsubsection{Mechanisms}
The mechanism assemblies required by LEMUR for operation of 
all subsystems and their expected characteristics and heritage are 
summarized in Table~\ref{table:mec}.
%%%%%%%%%%%%%%%%%%%%%%%%%%%%%%
% Table mechanisms
%%%%%%%%%%%%%%%%%%%%%%%%%%%%%%
\vspace{2ex}
\begin{table*}
\begin{center}
%%%%REMEMBER TO CHECK PAGINATION IN THE LAST VERSION
\caption[]{\label{table:mec}LEMUR mechanism table}
\begin{tabular}{p{2.3cm}p{4.8cm}p{3.5cm}}
\hline
\multicolumn{1}{c}{Mechanism} & \multicolumn{1}{c}{Requirements/characteristics} & \multicolumn{1}{c}{Heritage} \\
\hline & \mbox{}\\[-1.5ex]
Front door & Re-closable door mechanism with fail-safe redundant opening &  Secchi/HI on STEREO, SUMER and LASCO on SOHO \\[0.7ex]

Telescope pointing & Coarse: linear translation via dual motor stage along best image quality curve. \,~\,~\, Fine: two axis rotation for rastering and for image motion compensation using gimbal structure, flexural mounts, flex pivots, and voice coil drivers & Design similar to previous tip/tilt mirrors built by Ball Aerospace or similar to the SUMER pointing mechanism \\[0.7ex]

Slit select/focus & 5 slits on a linear translation mechanism  \,~\,~\,~\,\,\,\,~\,~\,~\,\,\,     $>$2~cm focus adjustment along the chief ray                   & SUMER/SOHO, EIS/Hinode \\[0.7ex]

Grating focus & $>$2~cm linear grating translation  & SUMER/SOHO, EIS/Hinode \\[0.7ex]

CCD shutter & Direct driven shutter mechanism with optical encoding. $>$4$\times10^{7}$ cycles  & EIS/Hinode, others\\[0.7ex]

ICCD detector door (optional) & One-shot door. Fail-safe actuator. Mechanical GSE for door re-closure during ground operations & HST and other missions \\[0.7ex]

\hline %%& \mbox{}\\[-1.5ex]

\end{tabular}
\end{center}
\end{table*}
%%%%%%%%%%%%%%%%%%%%%%%%%%%%%%%%%%%%%%%%%%%%%%%%%%%%%%%%%%%%%%%%%%%%%%%%%%%%	
\subsection{Radiometric Performance assessment}

\begin{figure}[t!]
\includegraphics[angle=0,width=120mm]{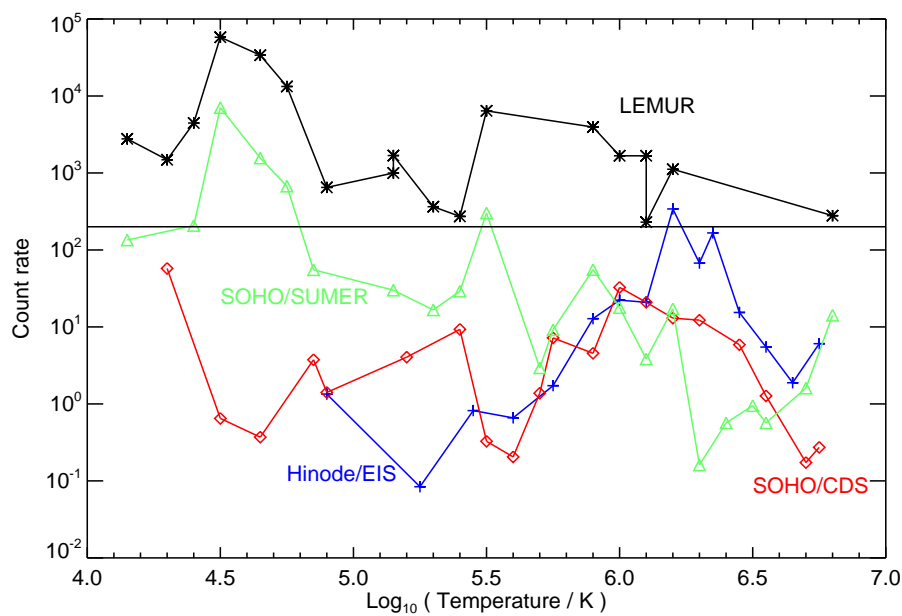} %%PS
\vskip -0.2cm
\caption[]{\label{fig:rad-perf} Comparison of the LEMUR expected 
count-rates with those from CDS and SUMER on SOHO and EIS on Hinode. 
The 200 count/s level is indicated by the solid horizontal line.}
\end{figure}

The two element optical design of LEMUR (minimum number of reflections), 
the large aperture, the good reflectance of the optics and the high quantum efficiency 
of the detectors, all contribute to an unprecedented throughput. 
Figure~\ref{fig:rad-perf} compares the count~s$^{-1}$ obtained from a solar
active region by LEMUR with 
those from the currently available spectrometers on Hinode and SOHO. 
Considering that about 200 counts in a spectral line are needed to measure
flows with a $<$2~km~s$^{-1}$ accuracy, it is evident that only LEMUR can and 
will obtain spectra of sufficient quality at all temperatures to perform the 
studies defined in Table~\ref{scitrace} and essential to address the science goals 
illustrated in Section~\ref{sci-obj-req-solarc}.

%%%%%%%%%%%%%%%%%%%%%%%%%%%%% 
% FIGURE radiometric performances
%%%%%%%%%%%%%%%%%%%%%%%%%%%%% 
%%%%%%%%%%%%%%%%%%%%%%%%%%%%%

%%%%%%%%%%%%%%%%%%%%%%%%%%%%%%%%%%%%%%%%%%%%%%%%%%%%%%%%%%%%%%%%%%%%%%%%%%%%	
\subsection{Summary of LEMUR nominal instrument resources}

A preliminary mass breakdown gives an estimated mass of 155~kg (15\% margin 
included) for LEMUR (85~kg for the telescope unit and 70~kg for the spectrograph 
unit). An estimate of the average dissipated power is of 68~W (including 24~W for 
operational heaters). Many of the resource allowances are derived from actual 
assemblies built for SERTS, EIS and SUMER.
The expected volume required for LEMUR is 430$\times$40$\times$70~cm$^3$. 
The predicted average data volume of LEMUR is of about 1.5~Mbps (after compression).

%%%%%%%%%%%%%%%%%%%%%%%%%%%%%%%%%%%%%%%%%%%%%%%%%%%%%%%%%%%%%%%%%%%%%%%%%%%%	
\subsection{Pointing and alignment requirements}

LEMUR will be precisely pointed to the Sun together with the Solar-C 
spacecraft.  During a single exposure, the pointing must be maintained 
to within $\sim$0.05\arcsec rms to maintain an overall resolution of 
0.28\arcsec.   The S/C Team at JAXA has conducted simulations of the spacecraft 
pointing performance with the Hinode pointing performance used as bench 
mark reference in these studies.  At the present time, the LEMUR line 
of sight is expected to be within the values given in Table~\ref{table:stab1}. 

%%%%%%%%%%%
% Table Axis stabilization
%%%%%%%%%%%%
\vspace{2ex}
\begin{table}
\begin{center}
\caption[]{\label{table:stab1}
Preliminary analysis results for LEMUR payload reference axis stabilization on the Solar-C S/C.}
\begin{tabular}{ll}
\hline
\multicolumn{1}{c}{Time period} & \multicolumn{1}{c}{expected stability} \\

\hline & \mbox{}\\[-1.5ex]

0.5~s & $<$0.3\arcsec  3 $\sigma$ \\
5~s & $<$0.3\arcsec 3 $\sigma$ \\
1~h & $<$2,0\arcsec  0 to peak \\
Mission life & $<$32.\arcsec  0 to peak \\
\hline %%& \mbox{}\\[-1.5ex]
\end{tabular}
\end{center}
\end{table}

%%%%%%%%%%%%%%%%%%%%%%%%%

The current S/C analysis indicates that LEMUR requires an active 
internal stabilization system. A comparison of the 
predicted performance of the LEMUR reference axis stability and the 
line of sight knowledge and stability requirements shows that 
modest line of sight correction is required 
within the LEMUR module. 

This modest correction can be achieved using the combination 
of a fine sun sensor (guide telescope), image motion stabilization 
(articulated primary mirror) and internal references (slit imaging 
assembly).  The primary control loop for line of sight 
stabilization is implemented using the output of the guide 
telescope to steer small corrective motions of the telescope 
mirror.  Initial calculations show that this control loop can
operate at $>$10 Hz frequency and provides the required 
0.15\arcsec (3$\sigma$) stabilization. Spacecraft jitter above 
the control 
frequencies are expected to be similar to SOT/Hinode. The SUVIT 
science instrument on the SOLAR-C payload requires a more stringent 
stability of 0.015\arcsec.  The SUVIT instrument also incorporates 
an internal motion stabilization system.

%%%%%%%%%%%%%%%%%%%%%%%%%	
\subsection{Special issues: radiometric calibration}
The LEMUR radiometric calibration concept uses the same 
approach successfully applied to calibrate the SOHO and 
Hinode VUV spectrographic instrumentation.  Component 
efficiencies will be characterized over their entire area 
using existing facilities at synchrotron radiation laboratories.  
Both detector and optical efficiencies will be measured.  The 
Telescope Unit itself will not need a full radiometric 
calibration. The end-to-end calibration of LEMUR (Telescope and 
Spectrograph Unit assembly) will be made using a VUV 
transfer source standard traceable to a primary source. 
Past experience demonstrates that 
an absolute radiometric calibration with uncertainties $<$15\% 
over the entire spectral range can be achieved.   
In-orbit flat-field response changes are monitored by placing 
the wide 5\arcsec slit in the focal plane and then moving the 
solar image at the slit in small increments while taking repeated 
exposures.  Similar techniques were used successfully to obtain 
the flat-field response of CDS-EIT-SUMER/SOHO, EIS/Hinode and 
TRACE.  Lastly, LEMUR will collaborate with sounding rocket 
flights to update the instrument calibration throughout its 
mission whenever possible.  A rigorous contamination control program 
will be implemented to maintain the instrument efficiency.

%%%%%%%%%%%%%%%%%%%%%%%%%%%%%%	
\subsection{Current heritage and Technology Readiness Level (TRL)}
\label{pay-trl}

LEMUR fully exploits extensive scientific and technical heritage in building 
large scale solar and spectroscopic instrumentation in Europe, the US and Japan. 
In fact, LEMUR is a larger scale version of instruments flown on previous missions (EIS 
aboard Hinode and SUMER aboard SOHO) and, as an entire instrument, is  overall at 
TRL 7. The assemblies and electronics boxes comprising LEMUR have extensive space 
flight heritage. 

%%%%%%%%%%%%%%%%%%%%%%%%%%%%	

\section{Summary and conclusions}
\label{sec-concl}
Our understanding of the outer solar atmosphere has been greatly improved 
by the successful solar missions launched in the past two decades. SOHO, Hinode, 
and SDO clearly show that the solar atmosphere is a very complex and dynamic 
environment where energy is transported and dissipated throughout all 
temperature regimes, with the magnetic field being the most important player.
To make a decisive step forward in our understanding of the solar (and hence 
stellar) magnetized atmosphere it is now necessary a comprehensive 
and simultaneous investigation of all the temperature regimes, from 0.01~MK up to 
the 20~MK observed during flares. These measurements must be as 
simultaneous as possible, have matching spatial resolutions, and need to be 
coupled to measurements of the magnetic field in the photosphere and 
chromosphere.
These are the drawing lines of the Japanese Solar C mission and its 
payload consisting of the three state-of-art instruments 
(here named SUVIT, XIT, and LEMUR), 
described in Section~\ref{sci-obj-req-solarc}.

LEMUR (Large European Module for solar Ultraviolet Research),
the subject of the present paper, provides the crucial link between 
the photospheric and chromospheric magnetic field and plasma
characteristics obtained by the visible telescope and
the high temporal and spatial resolution images of
the corona provided by the X-ray telescope.
Its principal science requirement is to
  obtain spectroscopic observations with sufficient resolution to
  measure the flow and dissipation of energy from the top layers of
  the chromosphere into the transition region and corona, and to
  observe multi-million-degree flare plasmas.  LEMUR will diagnose gas
  at all relevant temperatures with
  unprecedented spatial, spectral, and temporal resolution.  LEMUR's
  payload specifications are summarized in
  Table~\ref{table:solarc-payload}.

LEMUR consists of two major components: a
VUV solar telescope  with a 30-cm diameter
main mirror and a focal length of 3.6~m, 
and a focal-plane package
composed of VUV single-grating spectrometers
covering six carefully chosen wavelength ranges
between 170~\AA\ and 1270~\AA\ . 
The LEMUR slit covers 
280$\arcsec$ on the Sun with 0.14$\arcsec$ per 
pixel sampling. In
addition, LEMUR is capable of
achieving a temporal resolution of 0.5~s.

No major technology development is required to 
implement LEMUR.  It is at technology readiness 
level (TRL)~7 or higher for all components.

LEMUR has been proposed as a possible ESA-led 
contribution to the Solar C mission in the 2010 Call 
for M Class missions.
It will enable the European scientific 
community to participate as a key partner
in a mission that will revolutionize
our understanding of the magnetized
atmosphere of our own star.\\

\noindent 
{\small {\bf Acknowledgements} The work done by Naval Research 
Laboratory (NRL) personnel was supported by the NRL/ONR 6.1 program.}

% BibTeX users please use one of

\bibliography{LEMUR-paper}   % name your BibTeX data base

\bibliographystyle{spphys}       % APS-like style for physics

\end{document}